\documentclass{aa}  
\usepackage{tabularx}
\usepackage{booktabs}
\usepackage{array}
\usepackage{graphicx}
\usepackage{txfonts}
\usepackage{lipsum}
\usepackage{subcaption}
\usepackage{lscape}
\usepackage{placeins}
\usepackage{hyperref}

\begin{document}

   \title{High-resolution transmission spectroscopy of KELT-18b}

   \subtitle{Metastable helium detection at 1083 nm}

   \author{M. Chella\inst{1}\inst{2}
        \and M. Monticelli\inst{3}\inst{2}
        \and F. Biassoni\inst{2}
        \and F. Borsa\inst{2}
        \and M. Rainer\inst{2}
        \and G. Guilluy\inst{4}
        \and M. Basilicata\inst{2}
        \and L. Cabona\inst{2}
        }

   \institute{DiSAT, Università degli Studi dell’Insubria, via Valleggio 11, 22100 Como, Italy
   \and
   INAF - Osservatorio Astronomico di Brera, Via E. Bianchi 46, 23807 Merate (LC), Italy
   \and
    Department of Physics, Università di Genova, Via Dodecaneso 33, 16146 Genova, Italy
    \and
    INAF - Osservatorio Astrofisico di Torino, Via Osservatorio 20, 10025 Pino Torinese, Italy}

   \date{Received September 30, 20XX}

  \abstract
   {Exoplanets orbiting close to their host stars experience intense irradiation, which can drive the expansion and escape of their upper atmospheres. The near-infrared metastable helium triplet at 1083.3~nm has emerged as a powerful diagnostic to probe these mass-loss processes.}
   {We present the first high-resolution transmission spectroscopy of the ultra-hot Jupiter KELT-18b, probing its escaping atmosphere by analyzing a single transit obtained with the CARMENES spectrograph.}
   {Transmission spectroscopy and Doppler tomography were utilized to isolate the planetary signal. To mitigate the contamination from the strong telluric OH lines that overlap the region of interest, two distinct data analysis procedures were applied: a standard Molecfit and OH-masking approach, and a Principal Component Analysis (PCA) based method.}
   {By combining the two complementary approaches, we successfully detected a He\,I triplet absorption feature of $1.10 \pm 0.22\%$ ($5.1\sigma$). The signal exhibits a velocity shift compatible with zero and is attributed to the exoplanetary atmosphere. An injection test was necessary to reveal that the PCA-based method significantly suppresses the retrieved signal.
   Additionally, we modeled using the ATES and TPM codes the atmospheric photoevaporation and the helium triplet profile for different He/H ratios and stellar XUV fluxes, accounting for the instrument and planetary rotational broadening.
   From our simulations we derived a mass-loss rate and an evaporation efficiency for KELT-18b of $10.47 \leq \log_{10}(\dot{M}) \leq 11.68$ and $0.153 \leq \eta \leq 0.321$, showing that the atmosphere of KELT-18b is undergoing photoevaporation.
   Furthermore, our simulations show good agreement with the observed transmission spectrum, allowing us to exclude very low He/H ratios at a $3\sigma$ confidence level for XUV fluxes up to approximately twice the nominal reconstructed value.}
   {Our result expands the known parameter space of exoplanets with confirmed helium absorption, establishing KELT-18b as the most massive planet orbiting the hottest host star with such a detection. Furthermore, when employing PCA, we emphasize the necessity of performing injection tests to investigate whether any planetary signal is removed not only close to quadrature, but also during transits.}


   \maketitle
   \nolinenumbers

\section{Introduction}
\label{sec:intro}

The atmospheres of exoplanets orbiting close to their host stars can be strongly affected by intense stellar irradiation. In highly irradiated planets, the upper atmospheric layers can expand and, in some cases, undergo atmospheric escape.
High-energy stellar irradiation in the X-ray and extreme ultraviolet (XUV; $100 - 912$ \AA{}) bands plays a fundamental role in driving atmospheric escape. XUV photons deposit energy in the upper atmosphere, leading to heating, ionization, and hydrodynamic outflows for close in planets \citep{Lammer2003, Owen2012}. Over long timescales, this process can significantly alter planetary atmospheres and is thought to contribute to the observed scarcity of Neptune-sized planets at short orbital periods, known as the Neptune desert \citep[e.g.][]{Owen2018}.

In the last years the metastable helium triplet at $\sim$1083.3~nm has emerged as a powerful diagnostic for probing extended and escaping atmospheres. \cite{Seager2000} and later \cite{Oklopcic2018} identified the He\,I $2^{3}\mathrm{S}$ triplet as a particularly suitable tracer of atmospheric expansion and mass loss. Importantly, this feature lies in a region of the near-infrared spectrum that is largely unaffected by interstellar medium absorption and geocoronal emission, making it well suited for ground and space based observations.

The first detection of metastable helium in the atmosphere of an exoplanet was achieved using space-based observations with the Hubble Space Telescope (HST) targeting the warm Neptune WASP-107b \citep{Spake2018}. This achievement marked the beginning of the study of exoplanetary atmospheric evaporation using helium as a tracer, a key factor in understanding the evolution of exoplanetary systems. Ground-based high-resolution spectrographs have enabled similar detections \citep[e.g.][]{Nortmann2018,Allart2018,Salz2018}  establishing high-resolution spectroscopy as a powerful tool for studying atmospheric escape. High-resolution spectroscopy offers several key advantages for atmospheric characterization. By resolving individual spectral lines, it allows the separation of planetary signals from terrestrial and stellar contamination via their different Doppler shifts. Furthermore, it provides access to dynamic atmospheric information, such as wind patterns, which would be inaccessible at low spectral resolution \citep[e.g.][]{Snellen2010, Birkby2018}.

As the number of high-resolution atmospheric studies increases, with dozens of irradiated gas giants now surveyed for helium, clear demographic trends have begun to emerge \citep[e.g.][]{Allart2023, Orell-Miquel2026}. Both theoretical models and current observational data indicate that K-type stars provide the optimal high-energy irradiation environment to efficiently populate the metastable helium state \citep{Oklop2019}. Conversely, although hotter G-type stars exhibit a spectral energy distribution that is theoretically less favorable for sustaining a detectable helium population \citep{Oklop2019}, even hotter F-type stars present more favorable conditions, as demonstrated by successful helium detections up to a host star effective temperature of $\sim 6750$~K \citep{Orell-Miquel2026}. Although the theoretical mechanisms driving the metastable helium population in these hotter environments are still not fully understood, these high-temperature detections are strictly limited to highly irradiated planets with low bulk densities ($\rho_p \lesssim 0.5$~g \, cm$^{-3}$) that are likely to undergo intense and runaway atmospheric mass loss \citep{Thorngren2023}. Exploring these extreme boundary cases is essential to test the limits of these demographic trends and to fully understand the physical mechanisms driving atmospheric evaporation under extreme irradiation.

In this context, and across the growing number of high-resolution helium studies, the impact of telluric contamination and stellar variability remains a major unresolved challenge for ground-based observations.
In particular, telluric contamination represents a primary obstacle for ground-based high-resolution observations. In the near-infrared, and especially in the case of helium studies, the He\,I triplet region is heavily affected by strong and variable hydroxyl (OH) telluric emission lines \citep[e.g.][]{Allart2019, Guilluy2024}. Stellar variability further complicates the interpretation, as it is notoriously difficult to disentangle from the planetary signal \citep[e.g.][]{Cauley2018, Guilluy2020, Mercier2025}. One observational strategy to mitigate this issue consists in scheduling observations during transits where the combination of the target's systemic velocity and the Barycentric Earth Radial Velocity (BERV) shifts the telluric OH lines away from the expected planetary helium signal. Since scheduling observations during these optimal BERV windows is not always feasible, it is important to explore alternative approaches aimed at mitigating and directly removing the overlapping OH emission signal from the data.

In this work we search for helium absorption in the atmosphere of KELT-18\,b , an inflated hot Jupiter (see Table~\ref{table:1}) orbiting an F4V star on a polar orbit \citep{Rubenzahl2024}, with an equilibrium temperature of $\sim$ 2100~K and a low density ($\rho_{\rm p} = 0.377 \pm 0.040$ g\,cm$^{-3}$ , \citealt{McLeod2017}), conditions that make it a promising target for atmospheric escape studies \citep{McLeod2017, Orell-Miquel2026}.

This manuscript is organized as follows. In Sect.~\ref{sec:observations} we describes the observations and dataset. In Sect.~\ref{sec:data_analysis} we presents the data reduction and analysis methods used to extract the helium signal. The results are reported in Sect.~\ref{sec:results_validation}, followed by atmospheric escape modeling in Sect.~\ref{sec:modeling}. Finally, in Sect.~\ref{sec:conclusions} we discusses the implications of our findings.

\section{Observations}
\label{sec:observations}

In this study, we analyzed public archive data of KELT-18b obtained on 13 March 2018 with the CARMENES (Calar Alto high-Resolution search for M dwarfs with Exoearths with Near-infrared and optical Echelle Spectrographs) high-resolution spectrograph \citep{Quirrenbach2014}, mounted on the 3.5m telescope at the Calar Alto Observatory, Spain. The dataset was downloaded from the public archive of the Centro Astronómico Hispano en Andalucía (CAHA)\footnote{\url{http://caha.sdc.cab.inta-csic.es/calto}}. Specifically, we focused on data from the near-infrared channel (CARMENES-NIR), which covers a wavelength range from 0.96 to 1.71~$\mu$m with a spectral resolving power of $R \sim 80,400$. During the observations, fiber A was pointed at the target, while fiber B was illuminated by a calibration lamp. Consequently, fiber B could not be utilized to model and correct telluric emission \citep[e.g.,][]{Salz2018}. We utilized the reduced spectra provided by the standard CARMENES Data Reduction Software (DRS) pipeline.

\begin{figure}[ht!]
   \centering
   \includegraphics[width=\hsize]{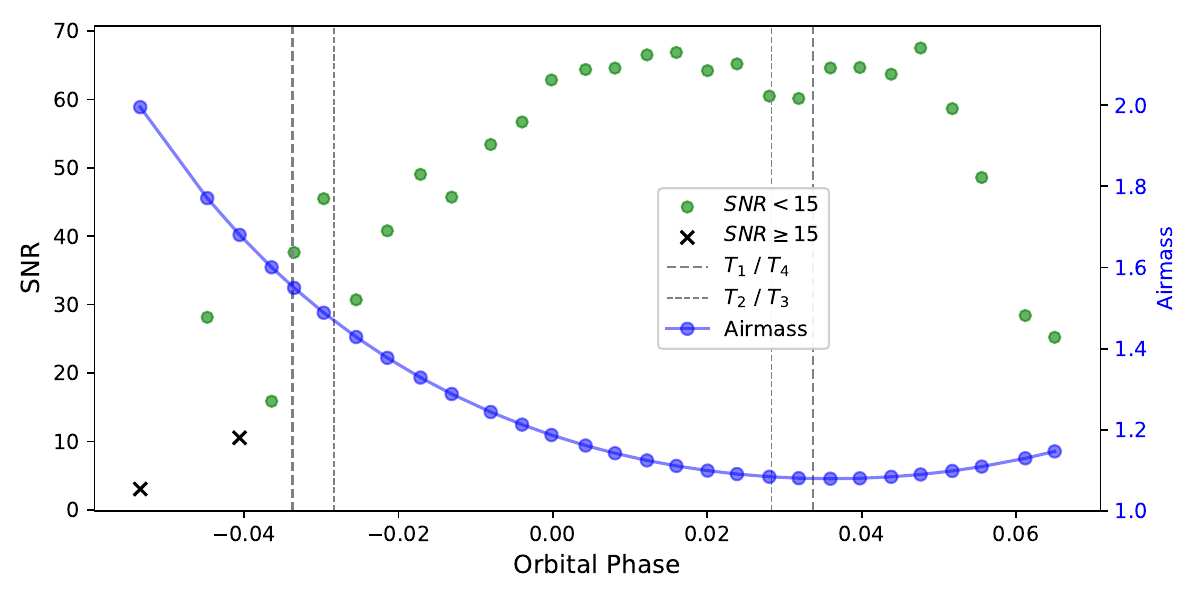}
      \caption{Evolution of the SNR and airmass during the observations. Green circles represent the data used in this work, while black crosses highlight the discarded exposures. The blue dots indicate the corresponding airmass values.}
         \label{fig:snr_airmass}
\end{figure}

The dataset is characterized by an average Signal-to-Noise Ratio (SNR) provided by the DRS pipeline of $\sim 49$ in spectral order 7, where the He\,I triplet is located. The target was observed at airmasses ranging from a minimum of 1.08 to a maximum of 2.00, with a mean value of 1.27. A complete log of the observations is reported in Table~\ref{table:obslog}. The evolution of both the SNR and the airmass as a function of the planetary orbital phase is shown in Fig.~\ref{fig:snr_airmass}. Prior to the data analysis, we inspected the time series and discarded two exposures (specifically, the first and third spectra, also highlighted in Fig.~\ref{fig:snr_airmass}) due to their low SNR. To ensure an adequate baseline coverage before the transit, we adopted a relatively loose quality threshold of $\text{SNR} = 15$. Retaining a sufficient number of pre-transit spectra is strictly necessary to properly model and remove the strong telluric contamination affecting the data, as shown in Sect.~\ref{sec:data_analysis}.

\begin{table}[ht!]
\caption{Stellar, planetary, and orbital parameters adopted in this work.}
\label{table:1}
\centering
\small
\renewcommand{\arraystretch}{1.4}

\begin{tabular}{l l r}
\hline\hline
Parameter & Value & Reference \\
\hline

\multicolumn{3}{l}{Stellar parameters} \\
Spectral type & F4V & \citealt{McLeod2017} \\
$T_{\rm eff}$ (K) & $6670 \pm 120$ & \citealt{McLeod2017} \\
$K_{\star}$ (m\,s$^{-1}$) & $127 \pm 11$ & \citealt{McLeod2017} \\
$M_{\star}$ ($M_{\odot}$) & $1.524^{+0.069}_{-0.068}$ & \citealt{McLeod2017} \\
$R_{\star}$ ($R_{\odot}$) & $1.908^{+0.042}_{-0.035}$ & \citealt{McLeod2017} \\
$\mathrm{[Fe/H]}$ (dex) & $0.090 \pm 0.130$ & \citealt{McLeod2017} \\
$v \sin i$ (km\,s$^{-1}$) & $10.2 \pm 1.0$ & \citealt{Rubenzahl2024} \\
$m_{\rm J}$ (mag) & $9.454 \pm 0.031$ & \citealt{McLeod2017} \\
$V_{\mathrm{sys}}$ (km\,s$^{-1}$) & $-11.6 \pm 0.1$ & \citealt{McLeod2017} \\
Age (Gyr) & $1.9 \pm 0.2 $ & \citealt{McLeod2017} \\

\hline
\multicolumn{3}{l}{Planetary parameters} \\
$M_{\rm p}$ ($M_{\rm J}$) & $1.18 \pm 0.11$ & \citealt{McLeod2017} \\
$R_{\rm p}$ ($R_{\rm J}$) & $1.570^{+0.042}_{-0.036}$ & \citealt{McLeod2017} \\
$T_{\rm eq}$ (K) & $2085^{+39}_{-38}$ & \citealt{McLeod2017} \\
$\rho_{\rm p}$ (g\,cm$^{-3}$) & $0.377 \pm 0.040$ & \citealt{McLeod2017} \\

\hline
\multicolumn{3}{l}{Orbital parameters} \\
$P$ (days) & $2.87169867 \pm 0.00000085$ & \citealt{Ivshina2022} \\
$T_0$ (BJD $_\mathrm{TDB}$) & $2458714.17773 \pm 0.00011$ & \citealt{Ivshina2022} \\
$i$ (deg) & $88.86^{+0.79}_{-1.20}$ & \citealt{McLeod2017} \\
$T_{14}$ (h) & $4.644^{+0.026}_{-0.024}$ & \citealt{McLeod2017} \\
$T_{12}$ (days) & $0.01545^{+0.00055}_{-0.00028}$ & \citealt{McLeod2017} \\
$a$ (AU) & $0.04550^{+0.00067}_{-0.00069}$ & \citealt{McLeod2017} \\
$e$ & 0 & \citealt{McLeod2017} \\
$b$ & $0.102^{+0.100}_{-0.071}$ & \citealt{McLeod2017} \\

\hline\hline
\end{tabular}
\end{table}

\begin{table*}[ht!]
\caption{Log of the observations.}
\label{table:obslog}
\centering
\renewcommand{\arraystretch}{1.3}
\begin{tabular}{l c c c c c c c}
\hline\hline
Date
& Instrument
& PI
& program n \tablefootmark{(a)}
& N$_{\rm obs}$ (in/out of transit)
& SNR\tablefootmark{(b)}
& Mean airmass
& $t_{\rm exp}$ (s) \\

\hline

2018-03-13 & CARMENES-NIR & Nascimbeni & 053 & 17/12 & 49 & 1.27 & 898s \\

\hline
\end{tabular}
\tablefoot{
\tablefoottext{a}{CAHA internal program number.}
\tablefoottext{b}{Time-averaged SNR of the order containing the He\,I triplet.}
}
\end{table*}

\section{Data analysis of the He\,I 1083.3 nm triplet}
\label{sec:data_analysis}

\begin{figure}[ht!]
   \centering
   \includegraphics[width=\hsize]{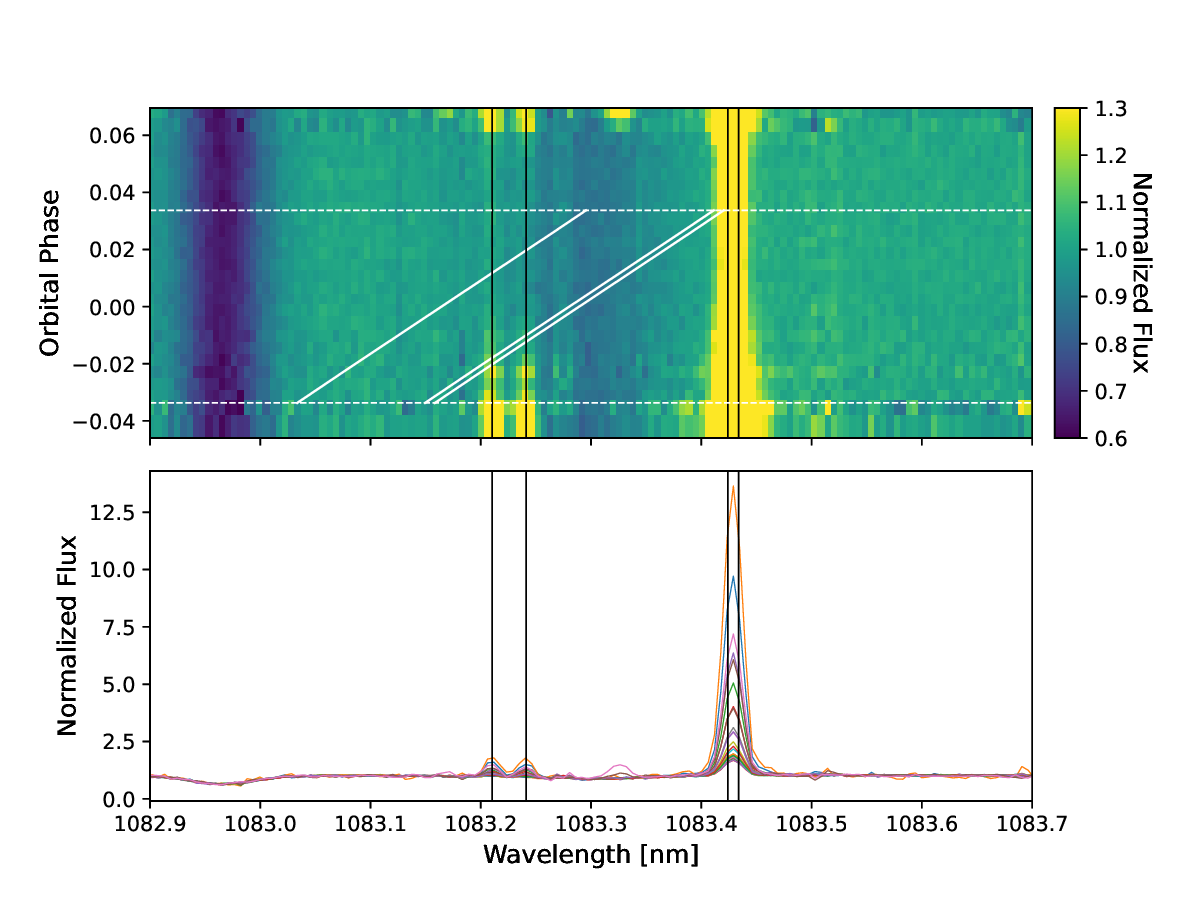}
      \caption{CARMENES spectra of KELT-18b in the He\,I triplet region, normalized for visualization. Top panel: flux as a function of wavelength and orbital phase. The white dashed lines indicate the start ($T_1$) and the end ($T_4$) of the planetary transit, while the solid white lines indicate the He\,I triplet in the planetary rest frame. Lower panel: single 1D spectra extracted by the CARMENES pipeline. The vertical black lines mark the vacuum wavelengths of the telluric OH lines. Note that the two OH lines at 1083.42 and 1083.43 nm appear blended due to the instrumental resolution.}
         \label{fig1}
\end{figure}

The metastable He\,I triplet, at 1083.206, 1083.322, and 1083.331~nm \footnote{We use vacuum wavelengths as a reference throughout this paper.}, lies in a spectral region frequently affected by strong telluric OH emission lines, which represent a well-known source of contamination in ground-based near-infrared observations \citep[e.g.,][]{Nortmann2018, Allart2019, Guilluy2020}. In this work, our goal is to extract the planetary absorption signal from this exact region which, as shown in Fig.~\ref{fig1}, is heavily impacted by such telluric features. The wavelengths of these OH lines, from the NIST database \citep{NIST_ASD}, are 1083.21, 1083.24, 1083.42, and 1083.43 nm. For specific combinations of systemic velocity and BERV, the expected planetary track at 1083.326~nm (computed as the average wavelength of the two strongest components of the He\,I triplet in the planet rest frame) falls between, and partially overlaps, with the telluric OH emission lines. To accurately extract the planetary signal, it is therefore essential to carefully remove this OH contamination.

In Sect.~\ref{subsec:masking} and~\ref{subsec:pca}, we illustrate the two different approaches used to analyze the He\,I 1083.3 nm triplet, which are two commonly used methods in the literature. The first approach faces telluric contamination in two distinct steps. Initially, a correction for telluric absorption lines is applied to the individual stellar spectra. Subsequently, the spectral regions contaminated by OH emissions are directly masked \citep[e.g.,][]{Guilluy2023}. The second method uses Principal Component Analysis (PCA) \citep{Kok2013} to simultaneously clean telluric emissions, telluric absorptions, and the stellar spectrum. While PCA is a standard technique in high-resolution transmission spectroscopy, it is not widely used specifically to investigate the He\,I triplet. However, several studies have successfully employed SYSREM \citep{Mazeh2007}, an algorithm similar to PCA, to analyze this specific spectral region \citep[e.g.,][]{Zhang2021, Zhang2022_63433c}. As we will show in this section, the PCA approach allows us to mask less planetary signal compared to the OH-masking method.

\subsection{Molecfit correction}
\label{subsec:molecfit}

The first method follows the standard transmission spectroscopy procedure outlined by \cite{wyttenbach2015}, introducing some main differences. Primarily, rather than relying on an empirical airmass-based correction, we utilized the Molecfit ESO software \citep{Smette2015,Kausch2015} as an initial step to model and remove telluric absorptions. Specifically, we accounted for H$_2$O, CO$_2$ and O$_2$, which are the main sources of telluric contamination in the He\,I triplet spectral order. We avoided the standard airmass-based correction because it could inadvertently alter or reduce the planetary signal if such a signal correlates with the airmass. Particularly for low-SNR datasets, using Molecfit ensures that the removed spectral features are strictly those predicted by physical telluric models.

\subsection{OH-masking approach}
\label{subsec:masking}

Once the absorptions are removed, we treated the telluric emissions. Our approach is to eliminate the spectral signal present in the OH emission regions by masking fluxes between 1083.197 and 1083.256 nm, and between 1083.397 and 1083.464 nm (as in \citealt{Guilluy2024}). After masking these data points, we resumed the standard \cite{wyttenbach2015} procedure.

First, the spectra were normalized using a sigma-clipped linear polynomial fit. To compute the local continuum, we selected two regions adjacent to the average center of the two deepest He\,I lines, specifically between $-20$~\AA\ and $-6$~\AA, and between $+6$~\AA\ and $+20$~\AA\,. Subsequently, the spectra were Doppler shifted into the stellar rest frame. This was done by defining the total velocity correction $\vec{v}_i$ for each exposure as the sum of the stellar Keplerian radial velocity ($\vec{v}_{\text{kep},\star}$), the systemic velocity ($\vec{v}_{\text{sys}}$), and the barycentric Earth radial velocity ($\vec{v}_{\text{BERV}}$), computed using the parameters listed in Table~\ref{table:1}:
$$ \vec{v}_i = \vec{v}_{\text{kep},\star} + \vec{V}_{\text{sys}} - \vec{v}_{\text{BERV}} $$

In this reference frame, each in-transit spectrum was divided by a master-out spectrum obtained by a weighted mean, based on pixel uncertainties, of all Out-Of-Transit (OOT) spectra. Then, the transmission spectrum was constructed by shifting the in-transit spectra into the planetary rest frame, accounting for the planetary Keplerian velocity, and combining them using a weighted average. We verified that computing the transmission spectrum using only the $T_2$--$T_3$ interval yields results fully consistent with the rest of our analysis. Finally, a local renormalization of the transmission spectrum was performed. This step is necessary to properly define the local continuum for the subsequent least squares fit of the Gaussian absorption profile. For this final local renormalization, the continuum was evaluated in the regions between $1082.900$~nm\ and $1083.226$~nm, and between $1083.426$~nm and $1083.700$~nm\ .

\subsection{PCA based approach}
\label{subsec:pca}

The second method employed to remove stellar and telluric effects from the data was Principal Component Analysis (PCA), following the procedure described by \citet{Pelletier2025a}. Briefly summarizing the key steps: after interpolating the spectra into the stellar rest frame, they were continuum-aligned using a boxcar smoothing and a Gaussian filter, following \citet{Gibson2022}. We noticed that applying the PCA in the telluric rest frame did not produce any significant difference, indicating that the choice between the stellar and telluric reference frames does not affect our results, as the two do not diverge significantly during the observation. Subsequently, a second-order polynomial fit of the median spectrum was divided out from all exposures, as in \citet{Brogi2019}. PCA was then applied to the resulting residuals in the wavelength domain. 

\begin{figure}[ht!]
   \centering
   \includegraphics[width=\hsize]{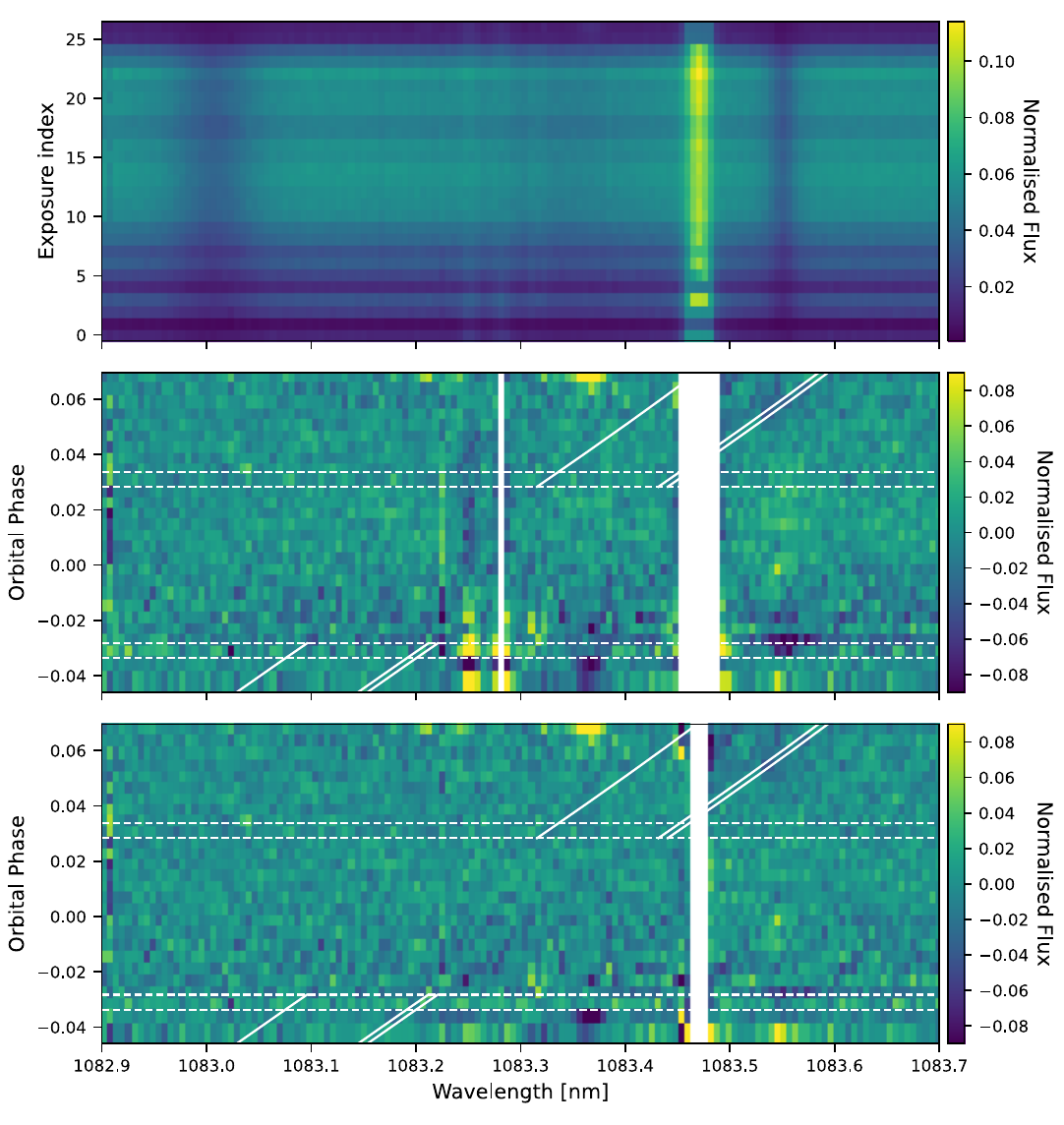}
      \caption{Data detrending result for the spectral order containing the Helium triplet. Top panel: 2D map of the spectra in the stellar rest frame before continuum alignment and normalization. Middle panel: Residuals after removing the first principal component. PCA is applied to the continuum aligned and normalized spectra. Lower panel: Residuals after removing the second principal component. The white vertical band indicates the masked noisy column corresponding to the variable OH emission. White slanted lines indicate the predicted planetary trace, while the horizontal dashed white lines indicate the four contact points ($T_1$, $T_2$, $T_3$, and $T_4$) of the planetary transit.}
\label{fig:pca_reduction}
\end{figure}

To avoid degrading the planetary signal, several tests were conducted with different numbers of removed PCA components. As empirically demonstrated in Sect.~\ref{subsec:res_oh_masking} and supported by our modeling in Sect.~\ref{sec:modeling}, the planetary He\,I absorption feature is intrinsically broad. When employing PCA, this characteristic is critical: due to its broadness, the planetary signal overlaps with itself across consecutive exposures, persisting in the same wavelength channels over time. As demonstrated by \citet{Pino2022}, PCA strongly suppresses features that appear stationary, specifically causing planetary signals near quadrature to be almost entirely removed by the detrending process. As the algorithm efficiently filters out time-correlated signatures, the removal of principal components inevitably degrades the underlying planetary signal, particularly for broad spectral features. Therefore, we identified the removal of two components as the optimal compromise for cleaning the spectra while preserving the planetary signal. The visual impact of this process on the spectral time-series is shown in Fig.~\ref{fig:pca_reduction}. It is clearly visible that, with just one component, significant telluric residues remain; these are significantly reduced when the second component is removed. Removing further components increases the loss of the planetary signal, which is already weakened by the two-component PCA process, resulting in a non-significant detection, as verified by injection tests in Sect.~\ref{subsec:injection}.

Following the PCA filtering, it was necessary to mask noisy columns exhibiting a standard deviation four times higher than the median of the spectral order. This final step is particularly critical for the helium analysis due to the highly variable OH emission lines at 1083.42~nm and 1083.43~nm, which cannot be fully removed by the PCA alone and a masking is needed. Finally, the uncertainties associated with the flux values for each pixel in this method were calculated using the analytical model presented by \citet{Gibson2020}. To calculate the transmission spectrum, we followed the same procedure described in Sect.~\ref{subsec:masking}.

\subsection{Validation of the expected $K_p$}
\label{subsec:mass_uncertainty}

\begin{figure}[ht!]
    \centering
    \includegraphics[width=\hsize]{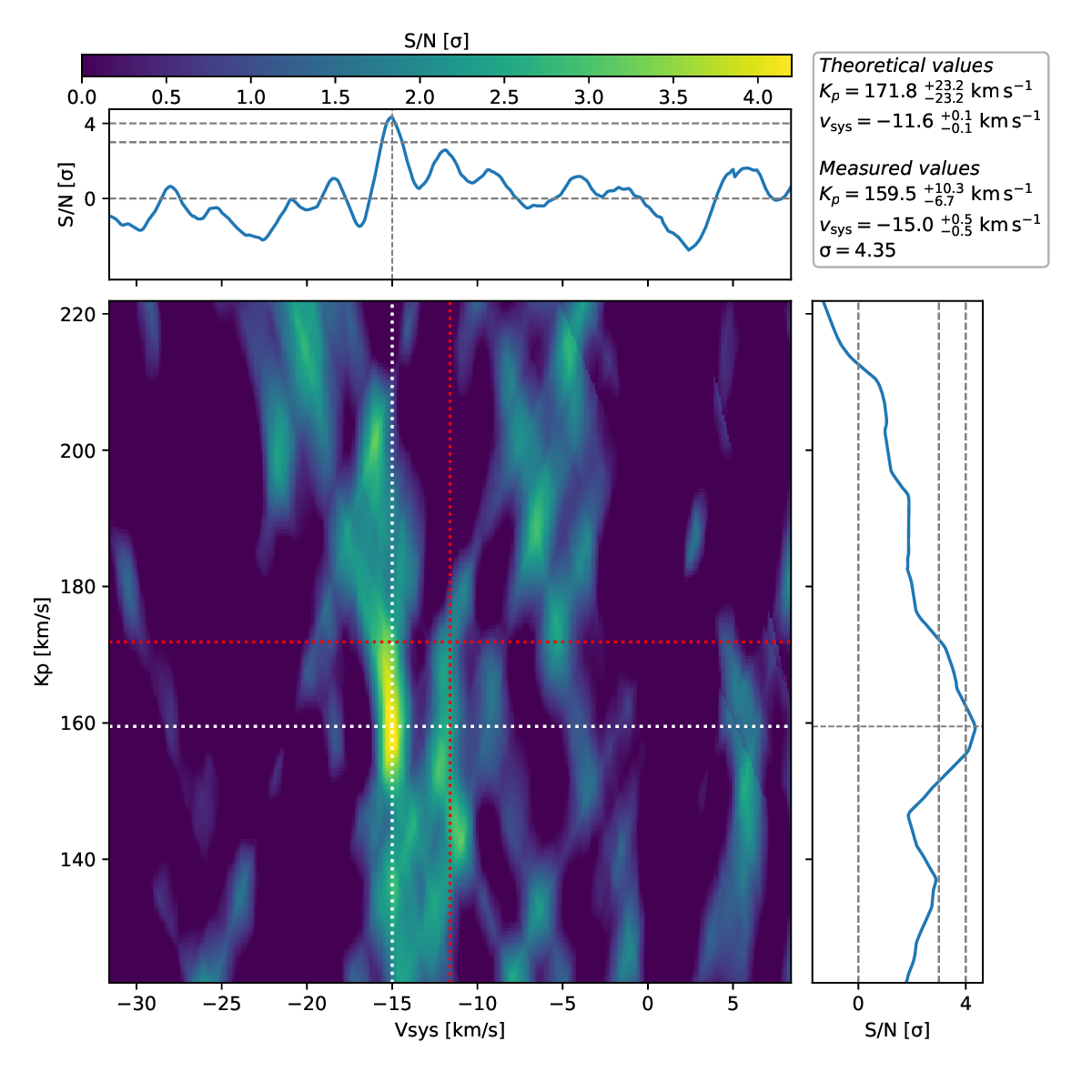}
    \caption{Signal-to-noise ratio (S/N) map in the $K_p$--$V_{\mathrm{sys}}$ parameter space, obtained by shifting and co-adding the transmission spectra over a grid of trial orbital semi-amplitudes $K_p$ and systemic velocities $V_{\mathrm{sys}}$. In the main 2D map, the white dotted lines mark the location of the measured peak, while the red dotted lines indicate the expected theoretical values. The top and right panels display 1D slices intersecting the peak along the $V_{\mathrm{sys}}$ and $K_p$ axes, respectively.}
    \label{fig:KpVsys}
\end{figure}

Given the $\sim 10\%$ uncertainty (see Table~\ref{table:1}) on the planetary mass, a rigorous approach requires evaluating the planetary absorption across a range of planetary radial velocity semi-amplitudes ($K_p$), since $K_p \propto 1/M_p$ (as $K_p/K_\star = M_\star/M_p$). To achieve this, we employed a $K_p - V_{sys}$ map, a well-established method in high-resolution cross-correlation atmospheric studies \citep[e.g.,][]{Birkby2018, Hoeijmakers2019}. Analogous to the cross-correlation procedure, we constructed this map by integrating the absorption signal along theoretical planetary tracks, computed for a grid of $K_p$ and systemic velocity values.  For the computation of this specific map, we utilized 4 PCA iterations (see Sect.~\ref{subsec:pca}) and raised the $\text{SNR}_{\text{threshold}} = 35$ (see Sect.~\ref{sec:observations}). The number of PCA components was selected following the statistical approach described by \citet[Fig.~1]{Basilicata2025}, which evaluates the relative variation of the residuals' standard deviation to identify when the telluric contamination is effectively removed. This more aggressive filtering was necessary to suppress spurious peaks in the $K_p - V_{\text{sys}}$ map caused by telluric residuals and to exclude the noisier spectra collected during the first part of the observation. Unlike the previous analysis, our primary objective here is not to preserve the true planetary absorption depth or the full trace in the Doppler tomography, but rather to evaluate at which $K_p$ the He\,I signal is maximized. Through this process, we obtain a peak in the map, visible in Fig.~\ref{fig:KpVsys}, corresponding to the maximum absorption signal at a specific $K_p$ and systemic velocity $V_{sys}$. In other words, by applying this procedure to our single-line analysis, we maximize the absorption in the transmission spectrum as a function of $V_{sys}$ and $K_p$, and consequently, of the planetary mass. As shown in Fig.~\ref{fig:KpVsys}, the map's peak at $K_p = 160.5^{+11.2}_{-7.7}$~km\,s$^{-1}$ is compatible within $1\sigma$ with the theoretical value of $K_p = 171.8 \pm 23.2$~km\,s$^{-1}$. We verified that adopting the empirical value of $K_p = 160.5$~km\,s$^{-1}$ yields results fully consistent with the rest of our analysis. Therefore, we chose to proceed by adopting the theoretical $K_p$ derived from the mass reported by \citet{McLeod2017}. Finally, we note a blueshift of $\sim 3.4$~km\,s$^{-1}$ between the velocity measured in the $K_p - V_{\mathrm{sys}}$ map ($-15.0 \pm 0.5$~km\,s$^{-1}$) and the theoretical value ($-11.6 \pm 0.1$~km\,s$^{-1}$, Table~\ref{table:1}), which is consistent with the presence of day-side to night-side winds \citep[e.g.,][]{wyttenbach2015}. 

\section{Results and validation}
\label{sec:results_validation}

\subsection{OH-masking approach}
\label{subsec:res_oh_masking}

\begin{figure}[ht!]
   \centering
   \includegraphics[width=\hsize]{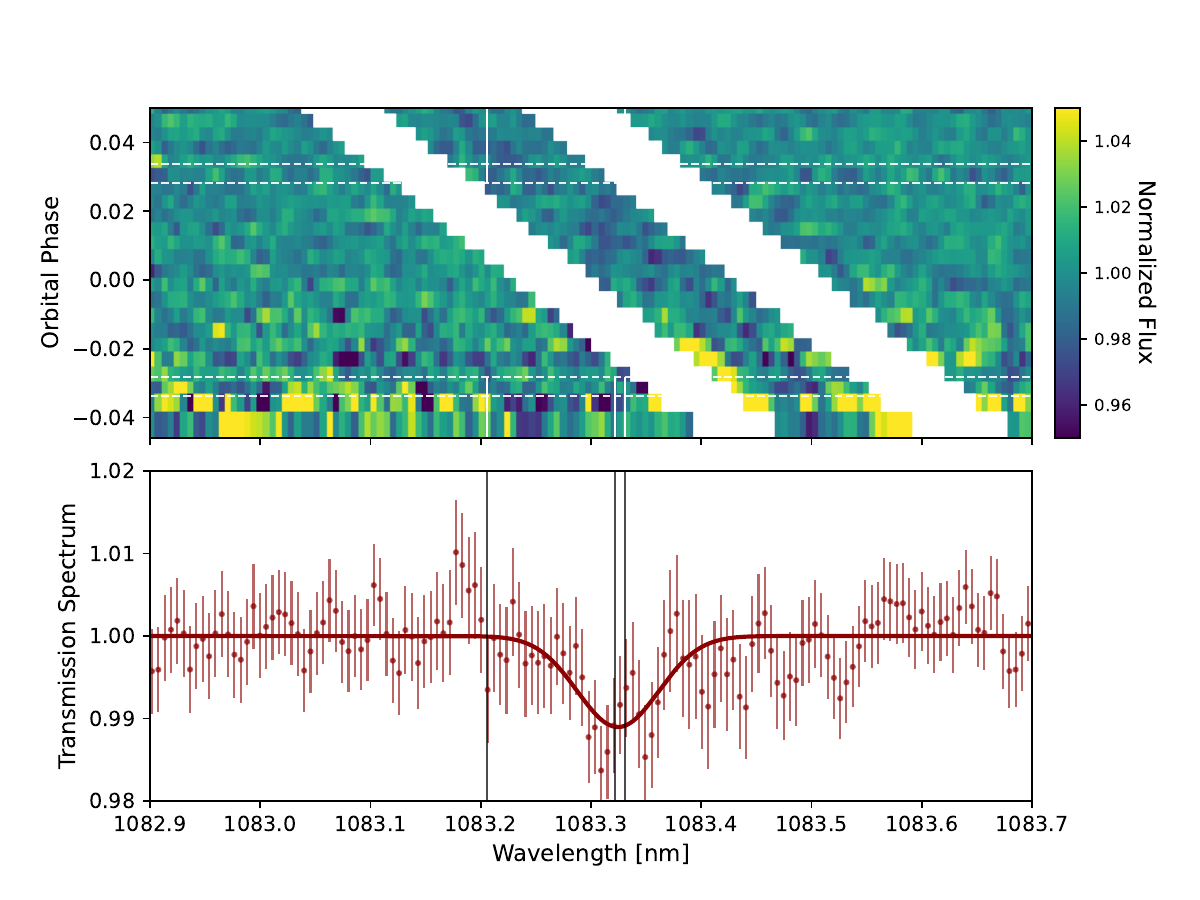}
      \caption{Analysis of the He\,I triplet region using the OH-masking method. The helium triplet in the planetary rest frame is indicated with vertical solid lines. Top panel: Doppler tomography aligned in the planetary frame. Lower panel: 1D transmission spectrum. The solid line represents the Gaussian fit of the absorption feature. Evident from the tomography is the emission signal to the right of the reddest OH line.}
    \label{fig:tomography}
\end{figure}

The transmission spectrum obtained with the OH-masking approach (Sect.~\ref{sec:data_analysis}) shows an absorption feature corresponding to the two deepest lines of the He\,I triplet in the planetary rest frame, as shown in Fig.~\ref{fig:tomography}. The absorption has a statistical significance of $5.1\sigma$, and the measured velocity shift is compatible with zero (see Table~\ref{tab:results}). 

The tomographic representation (top panel, Fig.~\ref{fig:tomography}) is routinely used to highlight the true reference frame of an absorption signal \citep[e.g.,][]{Borsa2018}.
However, in our case, due to the masking process, we removed $\sim 37\%$ of the planetary signal during the transit at the wavelength of the two main He\,I lines, as shown in Fig.~\ref{fig:tomography}. Consequently, we are unable to trace the planetary absorption track in the Doppler tomography. For this reason, we cannot unambiguously attribute the observed absorption to the planetary rest frame, and while we obtain a significant He\,I signal in the transmission spectrum, we cannot conclusively attribute it to the planetary atmosphere using the OH-masking method.

\begin{figure}[ht!]
   \centering
   \includegraphics[width=\columnwidth]{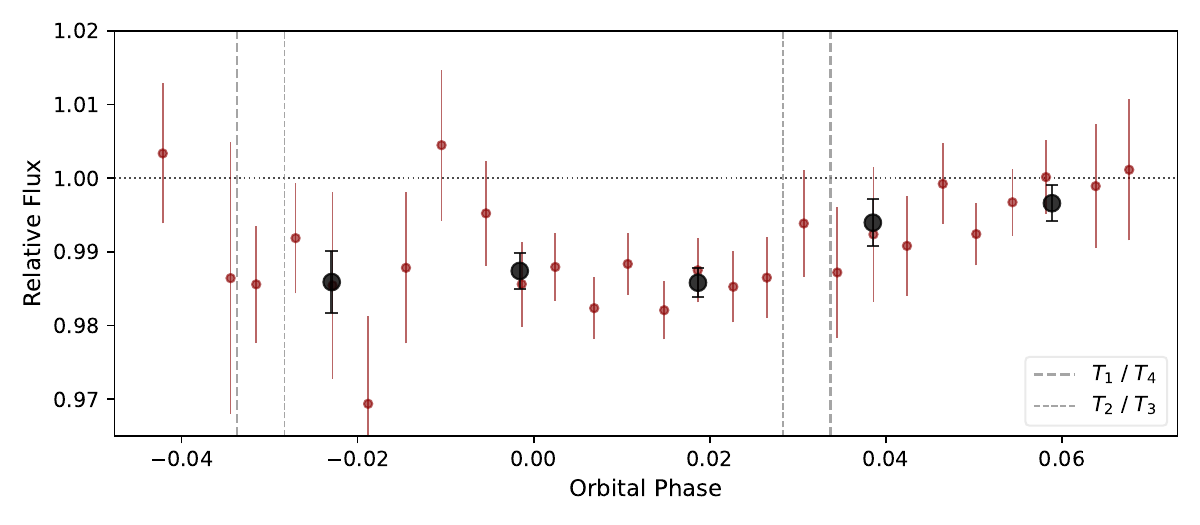}
    \caption{Spectrophotometric light curve integrated over a $1$\,\AA\ passband. The dark red points represent the original single exposures, while the black points indicate the data binned every five spectra, starting from the first in-transit exposure to avoid mixing out-of-transit and in-transit signal.}
   \label{fig:lightcurve}
\end{figure}

As in \citet{wyttenbach2015}, we studied the temporal evolution of the planetary absorption signal by extracting a spectrophotometric light curve. The curve, shown in Fig.~\ref{fig:lightcurve}, was obtained by integrating the signal within a $1$\,\AA\ passband, which corresponds to $27.7$\,km\,s$^{-1}$. The light curve hints at a possible post-transit absorption tail, however, as visible in Figs.~\ref{fig1} and \ref{fig:tomography}, during the egress phase the telluric OH emission lines overlap with the expected wavelength range of the planetary helium absorption. As also discussed in Sect.~\ref{subsec:photoevaporation_sim}, this contamination prevents a conclusive confirmation of a comet-like tail.

\begin{table}[ht!]
\caption{Fitted parameters for the He\,I triplet transmission spectrum using the two different reduction approaches.}
\centering
\renewcommand{\arraystretch}{1.3} 
\begin{tabular}{lcc}
\hline\hline
Parameter & OH masking & PCA \\
\hline
Amplitude (Depth) [\%] & $1.10 \pm 0.22$ & $0.63 \pm 0.18$\\
Significance [$\sigma$] & $5.1$ & $3.5$ \\
Velocity Shift [km\,s$^{-1}$] & $-0.5 \pm 2.4$ & $1.3 \pm 2.0$ \\
FWHM [km\,s$^{-1}$] & $24.0 \pm 5.9$ & $13.7 \pm 4.6$ \\
Equivalent Width [m\AA] & $10.2 \pm 3.2$ & $3.3 \pm 1.5$ \\
\hline
\end{tabular}
\label{tab:results}
\tablefoot{Parameters derived from a least squares Gaussian fit of the transmission spectrum around the He\,I triplet.}
\end{table}

\subsection{PCA based approach}

\begin{figure}[ht!]
   \centering
   \includegraphics[width=\hsize]{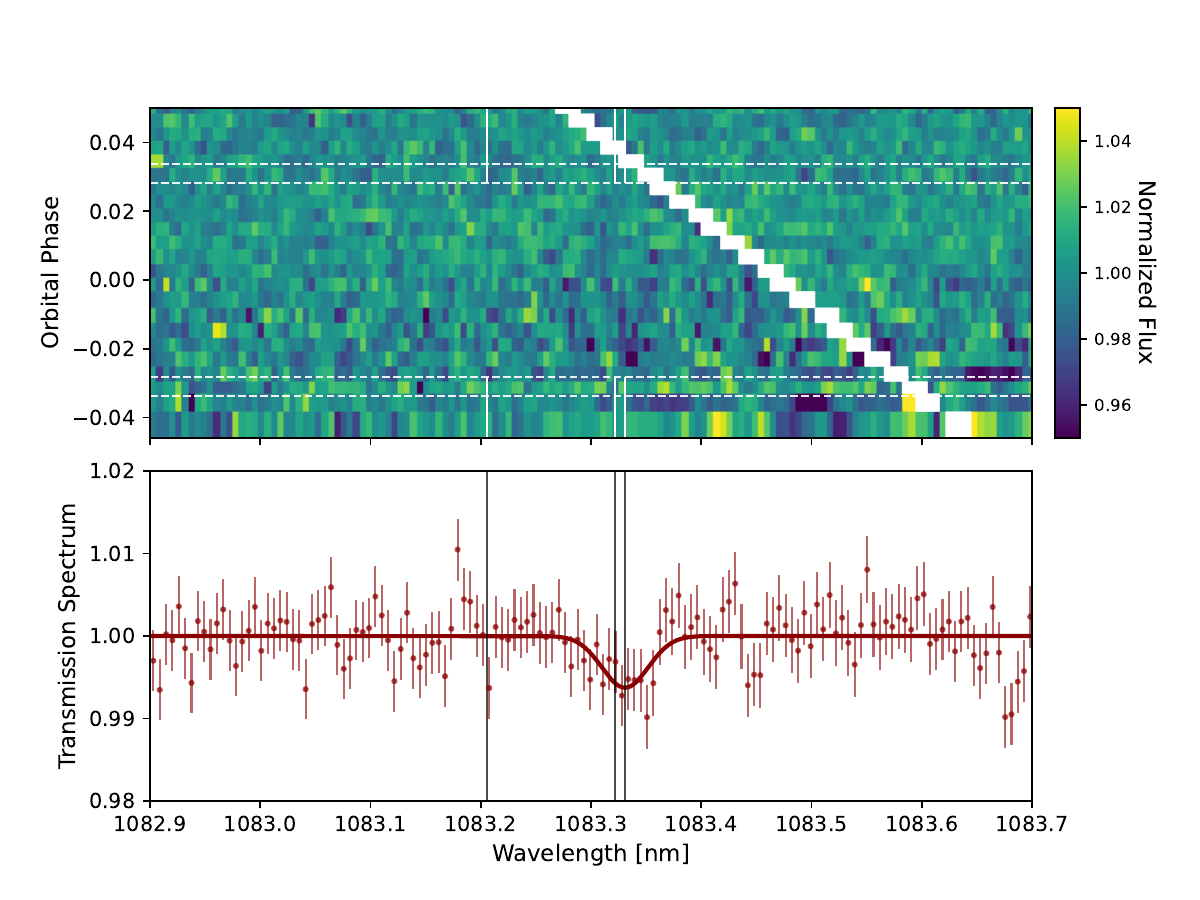}
      \caption{Analysis of the He\,I triplet region using the PCA-based approach. The helium triplet in the planetary rest frame is indicated with vertical solid lines. Top panel: Doppler tomography aligned in the planetary rest frame. Lower panel: 1D transmission spectrum. The solid line represents the Gaussian fit of the absorption feature.}
         \label{fig:tomography_pca}
\end{figure}

The transmission spectrum obtained with the PCA approach shows an absorption feature corresponding to the He\,I triplet with a statistical significance of $3.5\sigma$. Since the PCA technique allows us to minimize the masked spectral regions and mask less planetary signal, our final transmission spectrum is characterized by a higher SNR (see Table~\ref{tab:results}). We are now able to recover the planetary track in the Doppler tomography. By aligning the Doppler tomography in the planetary rest frame, an excess of absorption can be identified during the transit precisely at the rest wavelengths of the He\,I triplet coherent with the planetary restframe. As shown in Figures~\ref{fig:pca_reduction} and \ref{fig:tomography_pca}, this supports the association of the observed absorption with the exoplanet's atmosphere.

We note that only the Doppler shift parameter is comparable at $1\sigma$ between the two different approaches, as shown in Table~\ref{tab:results}. The PCA-based method exhibits a lower line depth, decreased significance, and a smaller FWHM, compared to the OH-masking approach. This behavior suggests that part of the planetary signal may be suppressed during the PCA detrending. 

Given their large uncertainties, the velocities in Table~\ref{tab:results} are compatible at $2\sigma$ with both the theoretical and measured systemic velocities reported in Sect.~\ref{subsec:mass_uncertainty}, although a difference is expected due to the two different analysis methods and $K_p$ values considered. Furthermore, to evaluate the dependence of the measured signal on the availability of OOT reference spectra, we tested the PCA detrending on exclusively pre-transit or post-transit baseline spectra. The retrieved planetary absorption depth remained consistent, indicating that the choice of the OOT reference spectra does not significantly bias the retrieved signal.

\subsection{Injection test}
\label{subsec:injection}

To test the latter hypothesis, an injection test was performed. We injected a Gaussian absorption signal, with parameters identical to those measured using the OH-masking approach (Sect.~\ref{subsec:res_oh_masking}, Table~\ref{tab:results}), into the CARMENES data. Specifically, the injection was performed just before the continuum alignment step defined in Sect.~\ref{subsec:pca}. The signal was injected into all in-transit spectra at a planetary rest frame wavelength of $1083.05$~nm, a region selected for its flat transmission spectrum. It was weighted across the different transit phases using a trapezoidal approximation \citep{Carter2008} i.e., by multiplying the injected signal by the photometric transit depth, to account for the reduced transmission signal present between $T_1$ and $T_2$, and between $T_3$ and $T_4$. Finally, to avoid artificially removing the injected signal, the local renormalization bands defined in Sect.~\ref{subsec:masking} were modified to exclude the region around 1083.05~nm.

\begin{table}[ht!]
\caption{Injected and retrieved parameters using the two different reduction approaches.}
\centering
\renewcommand{\arraystretch}{1.3}
\resizebox{\columnwidth}{!}{
\begin{tabular}{lccc}
\hline\hline
Parameter & Injected & OH masking & PCA \\
\hline
Amplitude (Depth) [\%] & $1.10$ & $0.96 \pm 0.19$ & $0.53 \pm 0.17$\\
Significance [$\sigma$] &  & $5.0$ & $3.2$ \\
Velocity Shift [km\,s$^{-1}$] & $0.0$ & $0.3 \pm 2.1$ & $-0.4 \pm 2.4$ \\
FWHM [km\,s$^{-1}$] & $24.0$ & $21.1 \pm 5.0$ & $15.3 \pm 5.6$ \\
Equivalent Width [m\AA] & $10.2$ & $7.8 \pm 2.4$ & $3.1 \pm 1.5$ \\
\hline
\end{tabular}
\label{tab:injection_results}
}
\tablefoot{The retrieved parameters are derived from a least-squares Gaussian fit.}

\end{table}

\begin{figure}[ht!]
   \centering
   \includegraphics[width=\hsize]{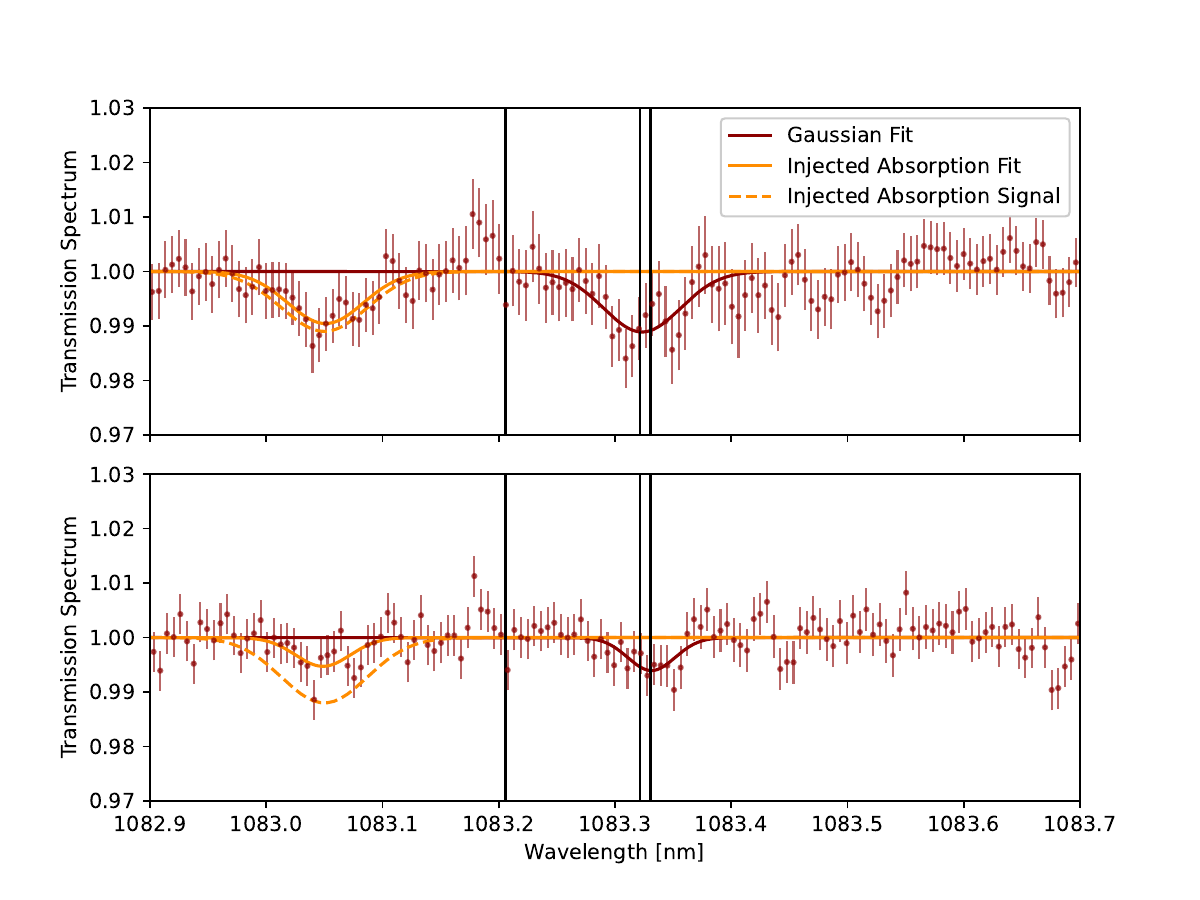}
      \caption{Transmission spectrum obtained by injecting the signal at 1083.05~nm. Top panel: Using the OH-masking method. Lower panel: Using the PCA-based approach.}
\label{fig:inj_comparison}
\end{figure}

\begin{figure}[ht!]
   \centering
   \includegraphics[width=\hsize]{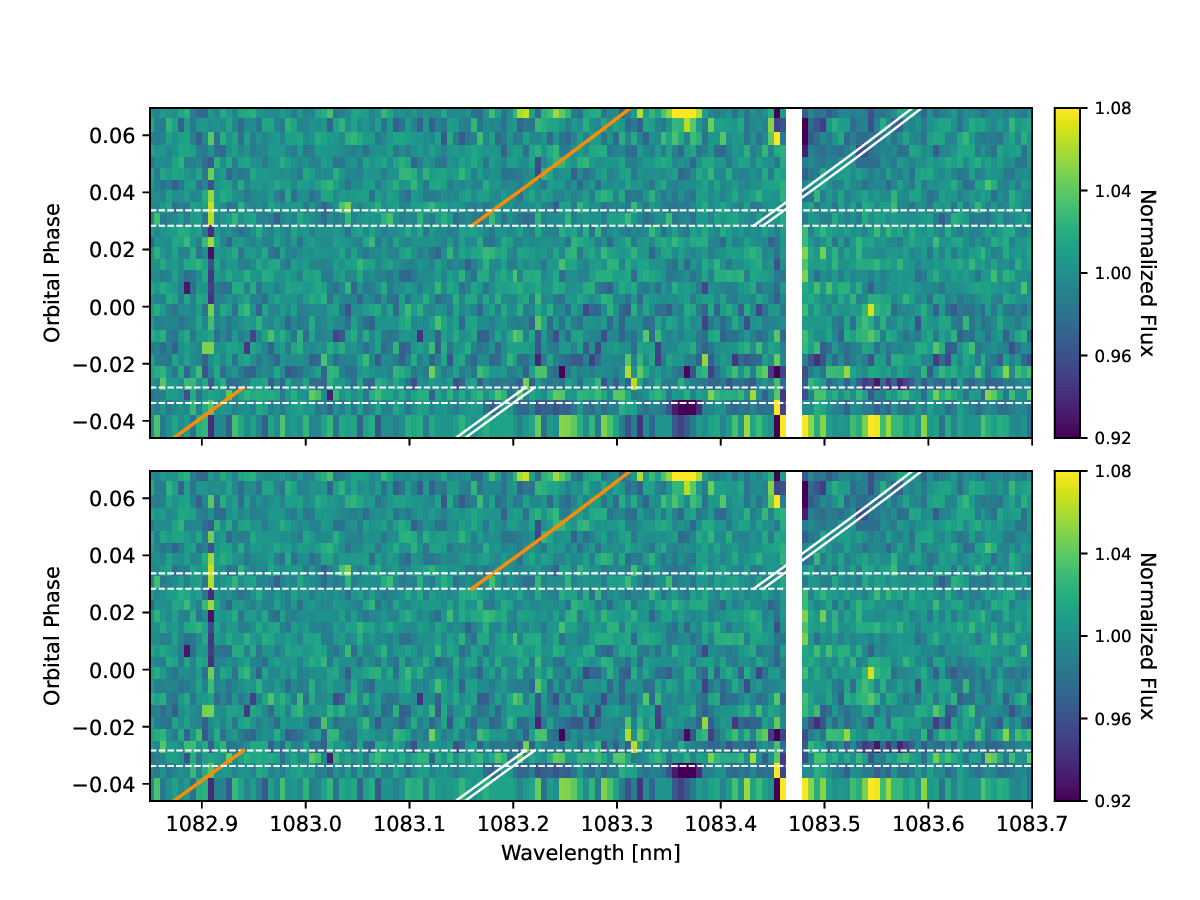}
      \caption{Doppler tomography of the PCA-based approach. The two white slanted lines indicate the two deepest lines of the He\,I triplet in the planetary rest frame, while the orange line indicates the injection wavelength. Top panel: Data with the injected planetary signal. Lower panel: Original data without injection.}
\label{fig:inj_doppler_tomography}
\end{figure}

As shown in Fig.~\ref{fig:inj_comparison} and Table~\ref{tab:injection_results}, the hypothesis that the PCA detrending partially removes the planetary signal is confirmed. While the PCA-based approach avoids masking spectral regions containing the planetary signal, it simultaneously removes $51.8\%$ of the absorption signal. A similar behavior was already observed for TOI 560.01 by \citet{Zhang2022_TOI560.01}, who reported a decrease of approximately half of the in-transit He\,I signal after SYSREM detrending. To investigate the impact of using more than two PCA components, we performed additional injection tests, which revealed that removing further components suppresses the retrieved signal below the $3\sigma$ significance threshold. For completeness, we also applied the injection test to the OH-masking method. In this case, the injected signal is fully recovered, as seen in the top panel of Fig.~\ref{fig:inj_comparison}. Furthermore, the Doppler tomography (Fig.~\ref{fig:inj_doppler_tomography}) reveals that the retrieved planetary signal and the injected one exhibit a comparable contrast.

\section{Atmospheric escape modeling}
\label{sec:modeling}

\subsection{Photoevaporation simulations}
\label{subsec:photoevaporation_sim}

We performed dedicated simulations to investigate the possible photoevaporation processes occurring in the atmosphere of KELT-18b. To this end, we made use of the publicly available code ATES \citep{Caldiroli_2021} \footnote{https://github.com/AndreaCaldiroli/ATES-Code}, adopting the planetary parameters (mass, radius, equilibrium temperature, and orbital distance), as well as the stellar mass given in Table~\ref{table:1}.
Other key inputs for these simulations are the stellar spectrum, particularly in the XUV wavelength range ($10 - 91.2$ nm), which plays a crucial role in the production of metastable helium, and the helium-to-hydrogen number density ratio (He/H) of the planetary outflow. Both stellar XUV spectrum and the planetary He/H are unknown parameters.
Observational constraints on stellar spectra in the XUV are generally lacking. For this reason, it is common practice to adopt reconstructed or proxy spectra from similar stars, especially for main-sequence, relatively quiescent stars, for which comparable spectral energy distributions are expected \citep[e.g.,][]{Biassoni2024}.
In this context, we adopted the MUSCLES synthetic spectrum \citep{Behr_2023} of WASP-17, which is a star consistent within $3\sigma$ with KELT-18 in terms of effective temperature ($T_{\rm eq.} = 6650 \pm 80$\,K), metallicity ([Fe/H]= $-0.19 \pm 0.09$), age ($2.65 \pm 0.25$\,Gyr) \citep{Bonomo_2017}, and rotational period ($P_{\rm rot} = 7.91 \pm 0.75$ days for WASP-17, \citealt{Anderson_2011} and $<8$ for KELT-18, \citealt{Rubenzahl2024}). The WASP-17 spectrum was then rescaled to the orbital distance of KELT-18b.
We used this spectrum as input for the ATES simulations, exploring five different values of He/H, a parameter that is not known a priori. According to literature results, atmospheric models are in agreement with observations assuming He/H values lower than $\sim 0.1$ \citep[e.g.,][]{Lampon_2020,Fossati2023}. Furthermore, the helium density may drop at the edge of the exoplanetary outflow due to diffusive processes \citep{Taylor_2025}. Hence, we considered for our simulations, a helium-to-hydrogen number density ratios of He/H = 0.01, 0.02, 0.05, 1/12 (cosmological value), and 0.10.

From these simulations, we derived mass-loss rates in units of [g/s] equal to 
$\log_{10}({\dot{M}}) = 11.08, 11.05, 10.94, 10.83,$ and $10.77$ for He/H = 0.01, 0.02, 0.05, 1/12, and 0.10, respectively. These values are relatively large for planets with gravitational potentials $ \log_{10}{|\Phi|} = \log_{10}{\left | \frac{GM_p}{R_p} \right|} > 13$, as is the case of KELT-18b (i.e., $\log_{10}{|\Phi|} = 13.13 \pm 0.04$), which are at the edge of the energy-limited regime and the recombination-limited regime \citep{Salz_2016,Caldiroli_2022}. Indeed, for high gravitational potential planets, a dominant fraction of the absorbed stellar flux is re-emitted locally, primarily through Ly$\alpha$ and bremsstrahlung emission \citep{Caldiroli_2022}.

To further assess the photoevaporation efficiency of KELT-18b, we evaluated it by comparing the mass-loss rates derived from ATES with those predicted by the energy-limited approximation given by:
\begin{equation}
    \dot{M} = \eta \frac{3 F_{\rm XUV}}{4 G K \rho_p}
    \label{eq.energy_limited}
\end{equation}

\noindent
where $G$ is the gravitational constant, $\rho_p$ is the mean planetary mass density and $K < 1$ is the potential energy reduction factor that accounts for the host star gravitational pull \citep{Erkaev_2007}. The $F_{\rm XUV}$ is the stellar flux at the planet distance in the XUV band, that for our spectrum used is $\log_{10}{F_{\rm XUV}} = 3.869$ where $F_{\rm XUV}$ has the units of [$erg \,\, cm^{-2} \,\, s^{-1}$]. The potential reduction factor $K$ can be expressed as $K = 1 - \frac{3}{2\xi} + \frac{1}{2\xi^3}$ \citep{Erkaev_2007}, where $\xi$ is the ratio between the planetary Roche radius and the planetary radius i.e., $\xi = R_{\rm RL} / R_p$, with the Roche radius $R_{\rm RL} = 3.89 R_p$ for KELT-18b.
To obtain the efficiency $\eta$ parameter we divided the mass-loss-rate $\dot{M}$ obtained with the ATES simulations by Eq.~\ref{eq.energy_limited} assuming for it $\eta = 1$.  
The resulting efficiencies are $\eta =$ 0.34, 0.32, 0.25, 0.19, and 0.17 for He/H = 0.01, 0.02, 0.05, 1/12, and 0.10 respectively. Despite its relatively high gravitational potential, the planet exhibits a substantial evaporation efficiency for low He/H values when compared to the energy-limited regime. Furthermore, the consistently high mass-loss rates derived from all five simulations indicate that the planet is undergoing active photoevaporation. Given its gravitational potential, KELT-18b sits exactly at the transition threshold between the energy-limited and recombination-limited regime \citep{Salz_2016}. Hence, different initial conditions as He/H, shift KELT-18b between these two evaporating regimes, increasing or decreasing the efficiency parameter, even at high values.

A more definitive confirmation of an ongoing photoevaporative process could be obtained by detecting a comet-like tail trailing the planet, extending the metastable helium absorption beyond the transit phase as observed in other systems such as WASP-69b and WASP-107b \citep{Nortmann2018, Spake2021}.
However, although the light curve hints at a possible post-transit absorption (Fig.~\ref{fig:lightcurve}), the presence of strong OH emission lines hampers its conclusive confirmation, as they overlap with the expected wavelength range of the planetary helium absorption at the end of the transit. Future observations, specifically scheduled to avoid contamination from OH emission by optimizing the transit timing relative to these features, will be crucial to conclusively confirm this scenario.

The ATES results of all the five simulations are depicted in Fig.~\ref{fig:ates_result}, from panel A to F. The overall outflow density, temperature and velocity are represented in panels A, B and C. The temperature profile increase sharply for all the simulations producing an inverted pressure-temperature profile up to $\sim 10^{-9}$ bar (panel D), where the temperature inverted the trend for the upper atmospheric layers and never resembles an isothermal profile. Due to the large gravitational potential, the outflow velocities are contained and always lower than 10 km s$^{-1}$ with the only exception of highest outflow layers. Hence, the broadening of the metastable helium triplet absorption profile is mainly due to the high temperature reached by the planetary outflow instead of escape velocity.

\subsection{Metastable helium modeling}

We used the Transmission Probability Module \citep[TPM;][]{Biassoni2024} to simulate the metastable helium triplet absorption profile using as input the velocity, temperature, and He\,I\,(2$^3$S) density profile obtained from the ATES' simulations, together with the stellar radius taken from Table \ref{table:1}. Panel G of Fig.~\ref{fig:ates_result} shows the TPM results for different He/H ratios. Since the temperature and velocity profiles of the escaping outflow remain closely consistent across all simulations, the primary difference between the absorption profiles is the assumed He/H ratio. Even though the local He\,I triplet number density does not reach extremely high peak values in the inner layers, its profile exhibits a gradual, slow decay rather than a sharp truncation, extending outward to large distances (Fig.~\ref{fig:ates_result}, panel F). This broad spatial extension ensures a substantial integrated column density along the line of sight during transit. Consequently, the resulting synthetic transmission spectra preserve a highly consistent line shape across all models, with the absorption depth scaling systematically with the total helium abundance, where the absorption peak ranges from 1.18\% to 0.39\% for He/H = 0.10 and 0.01, respectively.

Given the low hardness ratio of the host star, defined as the ratio between the XUV and FUV (91.2–258.3~nm) luminosities, $HR \sim 2.73 \times 10^{-4}$, one would expect a reduced population of metastable helium. Indeed, the relatively strong FUV flux efficiently depopulates the metastable state through photoionization (with the ionization edge at 258.3~nm), while the low XUV flux limits its production via ionization and subsequent recombination into the triplet state \citep{Biassoni2024}.
In addition, the high temperatures reached by the planetary outflow can further deplete metastable helium, as they enhance collisional processes with free electrons, one of the primary depopulation mechanisms alongside FUV photoionization \citep[Figure 1 of][]{Biassoni2024}.
For these reasons, hot stars of spectral type F and G are generally expected to be less favorable hosts for detectable metastable helium absorption, compared to K and M stars. The latter benefit not only from a more favorable hardness ratio, but also from geometric effects, that is a larger ratio between the atmospheric extent and the stellar radius, which increases the effective transit absorption signal \citep{Biassoni2024}.
However, recent studies have shown a growing number of detections around F-type stars, including systems similar to KELT-18 \citep{OrellMiquel_2024}, suggesting that this simple picture may be incomplete.
Our detection and theoretical models are fully consistent with this emerging trend, indicating that significant metastable helium absorption can still be present even in environments characterized by low XUV flux and strong FUV irradiation. This supports the idea that additional factors such as atmospheric structure, cooling and heating mechanisms provided by high or low metallicity, and high mass-loss-rate inducted by high bolometric fluxes of early type stars play a crucial role in regulating the population of metastable helium in the exoplanetary atmospheres.

\subsection{He/H -- $F_{\text{XUV}}$ degeneracy}
\label{subsec:quantitative_comparison}

In addition to the unconstrained He/H number density ratio, the exact stellar flux in the XUV band remains a major source of uncertainty. The WASP-17 XUV spectrum was derived by \cite{Behr_2023} according to empirical scaling relations provided by \cite{Linsky_2014}. These relations are affected by uncertainties of $\sim50\,\%$ \citep{Linsky_2014} in the XUV band.
For this reason, to account for these uncertainties, we used ATES and performed additional simulations by scaling the baseline XUV flux ($F_{0,\,\text{XUV}}$) multiplying it by factors of $\times 0.5\,, \times 2$, and $\times 5$, obtaining XUV fluxes of $\log_{10}F_{\rm XUV} = 3.568 \,, \, \log_{10}F_{\rm XUV} = 4.170$, and $\log_{10}F_{\rm XUV} = 4.568$, respectively. 
The ATES simulations were performed for each of the rescaled spectra and changing the He/H values, using the same five He/H previously adopted in the simulations with the nominal flux. In this way, we obtained a comprehensive grid of 20 different simulations exploring the coupled effects of varying stellar irradiation and atmospheric composition. Finally, for each simulation we used TPM to derive the transmission spectrum $T_{\text{He/H},\ F_{\text{XUV}}} (\lambda)$ for each He/H and stellar XUV irradiation.

\begin{figure}[t]
   \centering
   \includegraphics[width=\hsize]{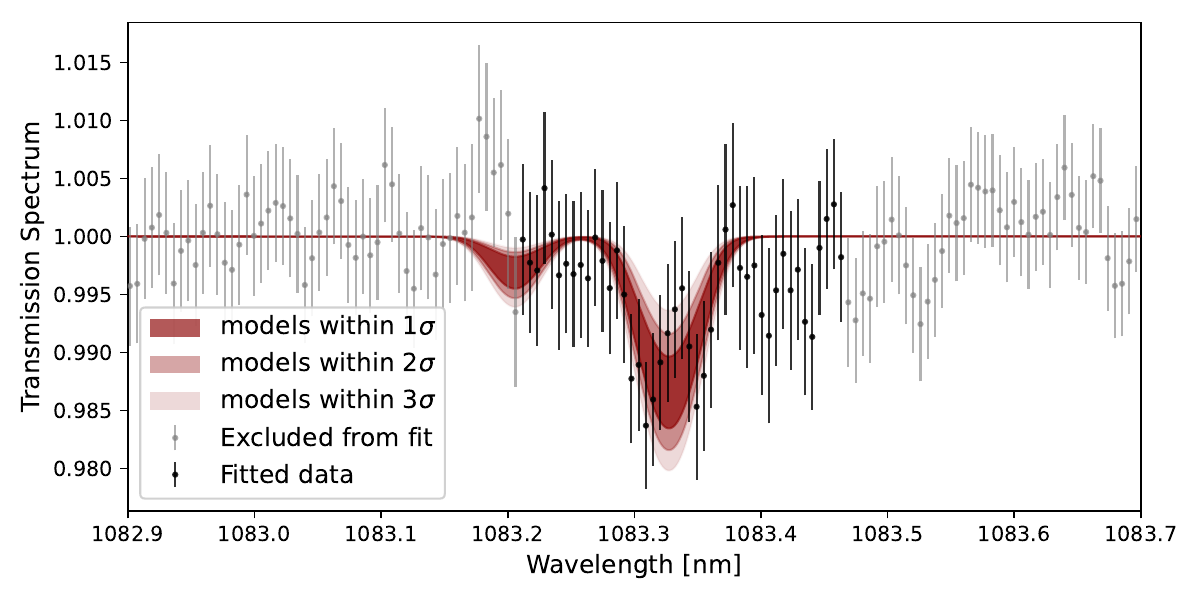}
   \caption{Transmission spectrum of KELT-18b around the metastable He\,I triplet. The shaded regions represent the envelopes of the models compatible at $1\sigma$, $2\sigma$, and $3\sigma$ confidence levels, derived from the $\chi^2$ map (Fig.~\ref{fig:chi2_map}).}
   \label{fig:models_overlapping}
\end{figure}

Starting from the 20 simulated transmission spectra $T_{\text{He/H},\ F_{\text{XUV}}} (\lambda)$, we performed a cubic interpolation of the simulated absorption profiles $A_{\text{He/H},\ F_{\text{XUV}}} (\lambda) = 1 - T_{\text{He/H},\ F_{\text{XUV}}} (\lambda)$ at each wavelength along the $\log_{10}(\text{He/H})$ and $\log_{10}(F_{\text{XUV}}/F_{0,\,\text{XUV}})$ parameter space over a $200 \times 200$ grid. For each grid point, we derived $T_{\text{He/H},\ F_{\text{XUV}}} (\lambda)$ from the interpolated $A_{\text{He/H},\ F_{\text{XUV}}} (\lambda)$ and calculated the $\chi^2$ value against the observed transmission spectrum. The fit was evaluated exclusively in the wavelength range between $1083.21$\,nm\ and $1083.47$\,nm, as shown in Fig.~\ref{fig:models_overlapping}. We excluded the shorter wavelengths from the computation because the continuum level in our models is fixed to 1. Furthermore, we excluded the data points around $1083.18$\,nm\ to avoid a likely telluric residual which partially overlaps with the weakest component of the triplet (it is clearly visible in the middle panel of Fig.~\ref{fig:pca_reduction} at $\sim 1083.22$\,nm\ in the telluric rest frame.)

\begin{figure}[t]
   \centering
   \includegraphics[width=\hsize]{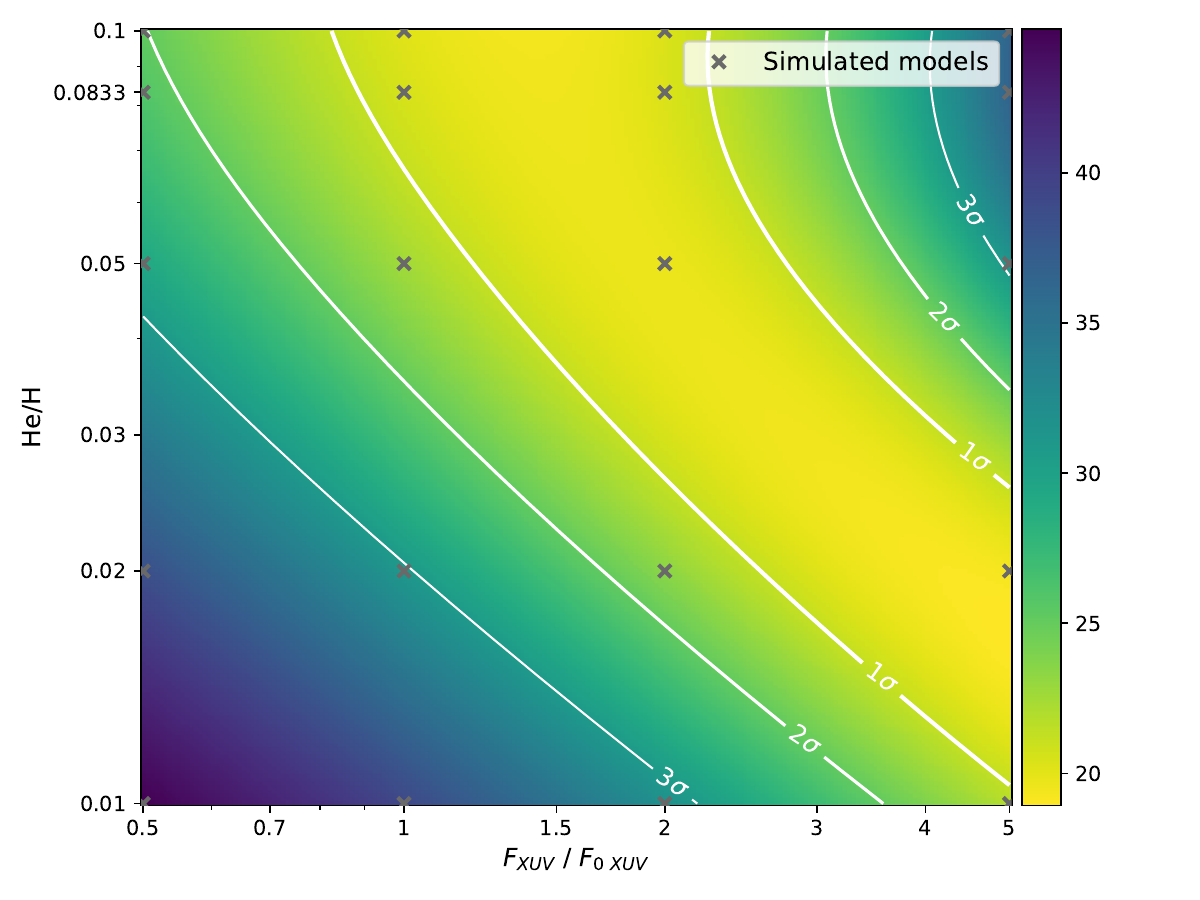}
   \caption{$\chi^2$ map of the explored parameter space obtained by comparing the interpolated theoretical models with the observed transmission spectrum. Contour lines represent the $1\sigma$, $2\sigma$, and $3\sigma$ confidence levels for the He/H and $F_{\text{XUV}}$ scaling factor.}
   \label{fig:chi2_map}
\end{figure}

The resulting $\chi^2$ map is presented in Fig.~\ref{fig:chi2_map}, displaying contour lines that indicate the significance levels at which different combinations of He/H and XUV flux can be excluded. To derive these contours, we computed $\Delta\chi^2 = \chi^2 - \chi^2_{\text{min}}$ for each point of the grid. Assuming a $\chi^2$ distribution with two degrees of freedom, the probability that the true model lies within a certain $\Delta\chi^2$ is given by $p = 1 - e^{-\Delta\chi_{k=2}^2/2}$. This probability was then converted into the equivalent standard deviations $\sigma$ of a 1D Gaussian distribution to define the confidence intervals. As shown in Fig.~\ref{fig:chi2_map}, there is a clear degeneracy between the He/H and the stellar XUV flux, a well-known behavior already discussed in the literature \citep[e.g.,][]{Biassoni2024}. Although our data do not allow us to retrieve a unique solution for these two parameters, we can still place robust constraints. Assuming XUV fluxes $\lesssim 2\,F_{0,\,\text{XUV}}$, we can exclude very low He/H ratios. The contour plot also illustrates that higher $F_{\text{XUV}}$ values are in agreement with progressively lower values of He/H. To visualize this agreement, the envelopes of the interpolated models falling within the $1\sigma$, $2\sigma$, and $3\sigma$ confidence intervals are overlaid on the measured transmission spectrum in Fig.~\ref{fig:models_overlapping}.
Figure \ref{fig:chi2_map} shows that lowering the XUV flux below the nominal value fails to yield any improvement in the overall fit quality. Consequently, further exploration towards even lower fluxes was deemed unnecessary, and the grid was restricted to the range spanning from $0.5\times$ to $5\times F_{\rm XUV}$, which effectively encompasses the representative uncertainties of the reconstruction models.

Using the same methodology, we interpolated the mass-loss rate $\dot{M}$ and the photoevaporation efficiency $\eta$ over the He/H $- F_{\text{XUV}}$  grid, producing the two additional maps presented in Appendix~\ref{sec:appendix_M_dot_eta}. Due to the observed degeneracy, we cannot establish strict values for $\eta$ and $\dot{M}$. Nevertheless, by evaluating these parameters within the $\chi^2$ confidence intervals derived above, we were able to constrain their allowed ranges. Specifically, assuming $0.5 \le F_{\text{XUV}}/F_{0,\text{XUV}} \le 5$ and $\text{He/H} \le 0.1$, the mass-loss rate is bounded between $\log_{10}(\dot{M}\,[g/s]) = 10.69$ and $11.68$ at the $1\sigma$ confidence level, and between $10.47$ and $11.68$ at $3\sigma$. Analogously, the efficiency is constrained between $\eta = 0.161$ and $0.284$ at $1\sigma$, and between $\eta = 0.153$ and $0.321$ at $3\sigma$.

To verify the robustness of our results, we also explored an MCMC approach and introduced a broadening parameter to account for atmospheric winds. These additional analyses confirm the observed degeneracy and the validity of the derived lower limits for the He/H and $F_{\text{XUV}}$. These alternative procedures and their limitations are discussed in Appendix~\ref{sec:appendix_x}.

Notably, unlike several previous studies comparing He\,I $(2^3\rm S)$ transmission spectra with theoretical models \citep[e.g.,][]{Fossati2022, Fossati2023, Lampon2023}, our measured absorption is compatible with the simulations even assuming a primordial He/H ratio. This agreement is quite surprising given the frequent discrepancies commonly reported in the literature for other planets, where theoretical models tend to overestimate the helium absorption depth unless anomalously low He/H ratios are adopted \citep[e.g.,][]{Kasper2020, Zhang2021, Fossati2022, Fossati2023, Lampon2023, Bennett2023, Vissapragada2024}.

We note that the reduced $\chi^2$ of our fit is less than one. This is likely due to an overestimation of the errors induced by the spectral interpolation, which also introduces correlations between adjacent pixels. Consequently, since our calculation assumes independent errors, the derived confidence contours might be optimistic.

\section{Discussion and conclusions}
\label{sec:conclusions}

\begin{figure*}[ht!]
    \centering
    \includegraphics[width=\hsize]{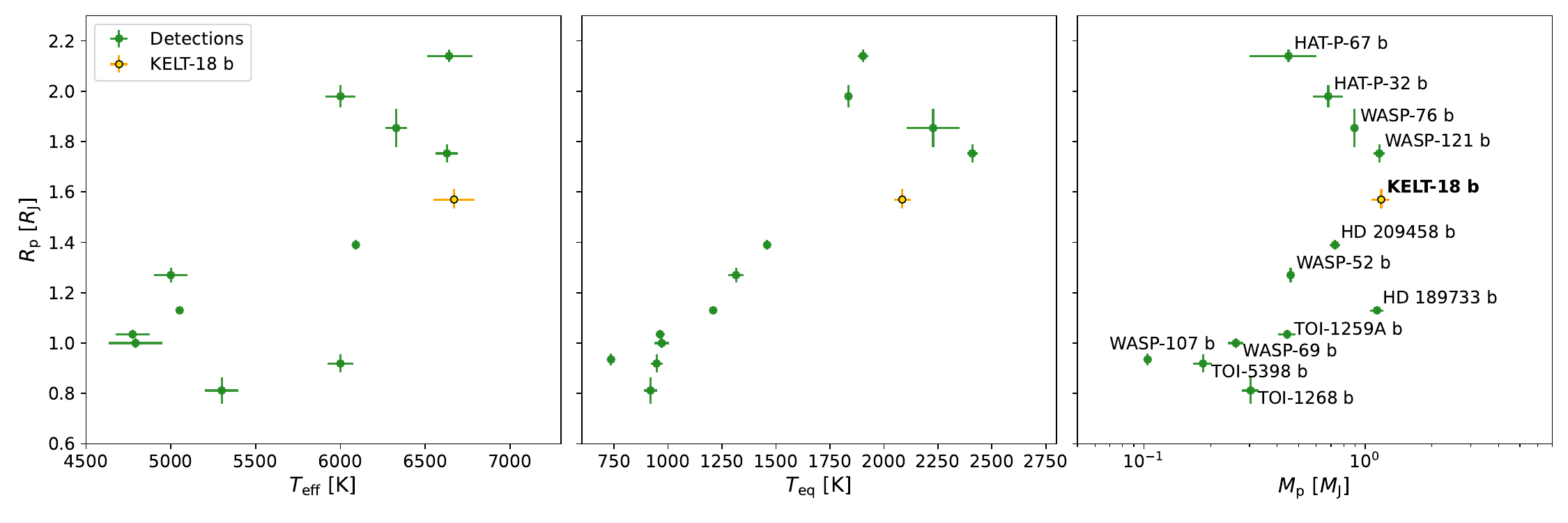}
    \caption{Distribution of He\,I triplet detections for gas giants ($R_p > 0.5~R_{\mathrm{Jup}}$) as a function of host star effective temperature ($T_{\mathrm{eff}}$, left panel), planet equilibrium temperature ($T_{\mathrm{eq}}$, central panel), and planetary mass ($M_{\mathrm{p}}$, right panel). Data were retrieved from the ExoAtmospheres database (IAC community database for exoplanet atmospheric observations) and the NASA Exoplanet Archive \citep{Christiansen2025}. Parameters for WASP-121b were adopted from \cite{Bourrier2020}.}
    \label{fig:detections_comparison}
\end{figure*}

In this work, we present the first high-resolution transmission spectroscopy of the ultra-hot Jupiter KELT-18b. Based on the analysis of the CARMENES-NIR dataset using two different data analysis approaches, we report an absorption signal of $1.1\%$ attributed to the He\,I triplet at $1083.3$~nm in the planetary atmosphere.

The specific observational conditions characterizing our dataset (i.e., the lack of sky emission monitoring from fiber B and the combination of systemic velocity, BERV, and planetary velocity semi-amplitude $K_p$ that placed the planetary track directly between the two OH doublets) presented a significant challenge for the signal extraction. However, we demonstrate that this limitation can be addressed through a combined application of the OH-masking and PCA-based techniques, which allows us to recover the absorption signal and supports its association with the exoplanetary atmosphere. Furthermore, in agreement with the existing literature \citep{Zhang2022_TOI560.01, Czesla2024_sysrem}, we confirm that the PCA method can significantly suppress the targeted planetary signal, depending on its strength and the $K_p$ value. Therefore, we highlight the importance of performing an injection test to monitor and quantify this effect.

From our ATES and TPM models, we demonstrate a good agreement between the simulated transmission spectra and our observational detection. Specifically, our quantitative analysis allows us to exclude very low He/H ratios at a $3\sigma$ confidence level, unless very high stellar XUV fluxes are assumed. Notably, this agreement between the theoretical models and the measured transmission spectrum is achieved without the need to assume anomalously low He/H ratios, avoiding the theoretical overestimations frequently reported in the literature for other exoplanetary systems. Given the high mass-loss rate and efficiency derived for KELT-18b i.e., $10.47 \leq \log_{10}(\dot{M}) \leq 11.68$ and $0.153 \leq \eta \leq 0.321$ at $3\sigma$ confidence level, we expect that the planet is undergoing atmospheric escape driven by photoevaporation. Future observations are necessary to confirm this hypothesis, specifically focusing on the detection of a comet-like tail. These observations should be carefully conducted under conditions where telluric OH lines do not overlap with the expected planetary signal.

Until now, the exoplanet orbiting the hottest host star with a confirmed helium detection was WASP-121b \citep{Czesla2024}, with a stellar $T_{\rm eff}$ of $6460$~K \citep{Delrez2016}. Our detection pushes this boundary further, placing KELT-18b at the upper extreme of the known parameter space for this type of atmospheric study (see Fig.~\ref{fig:detections_comparison}). Furthermore, KELT-18b ($M_p = 1.18~M_{\rm J}$) extends the currently explored planetary mass range, overcoming the $\sim 1.1~M_{\rm J}$ limit for previous helium detections reported by \citet{Orell-Miquel2026}. Consequently, KELT-18b sits just above WASP-121b in terms of host star temperature and planetary mass, although we note that these parameters remain compatible between the two systems when observational uncertainties are taken into account, as shown in Fig.~\ref{fig:detections_comparison}.

\begin{acknowledgements}
F.Bo. acknowledges support from Bando Ricerca Fondamentale INAF 2023 and INAF GO Large Grant 2024 "Strenghtening pylons for BRIDGES".
.
This research has made use of the Spanish Virtual Observatory (http://svo.cab.inta-csic.es) supported by the MINECO/FEDER through grant AyA2017-84089.7
Based on data from the CAHA Archive at CAB (INTA-CSIC). The CAHA Archive is part of the Spanish Virtual Observatory project funded by MCIN/AEI/10.13039/501100011033 through grant PID2023-146210NB-I00
\end{acknowledgements}

\bibliographystyle{bibtex/aa}
\bibliography{references}

\clearpage

\begin{appendix}

\onecolumn
\section{ATES simulations}

\begin{figure*}[ht!]
\centering
    \includegraphics[width=0.99\hsize]{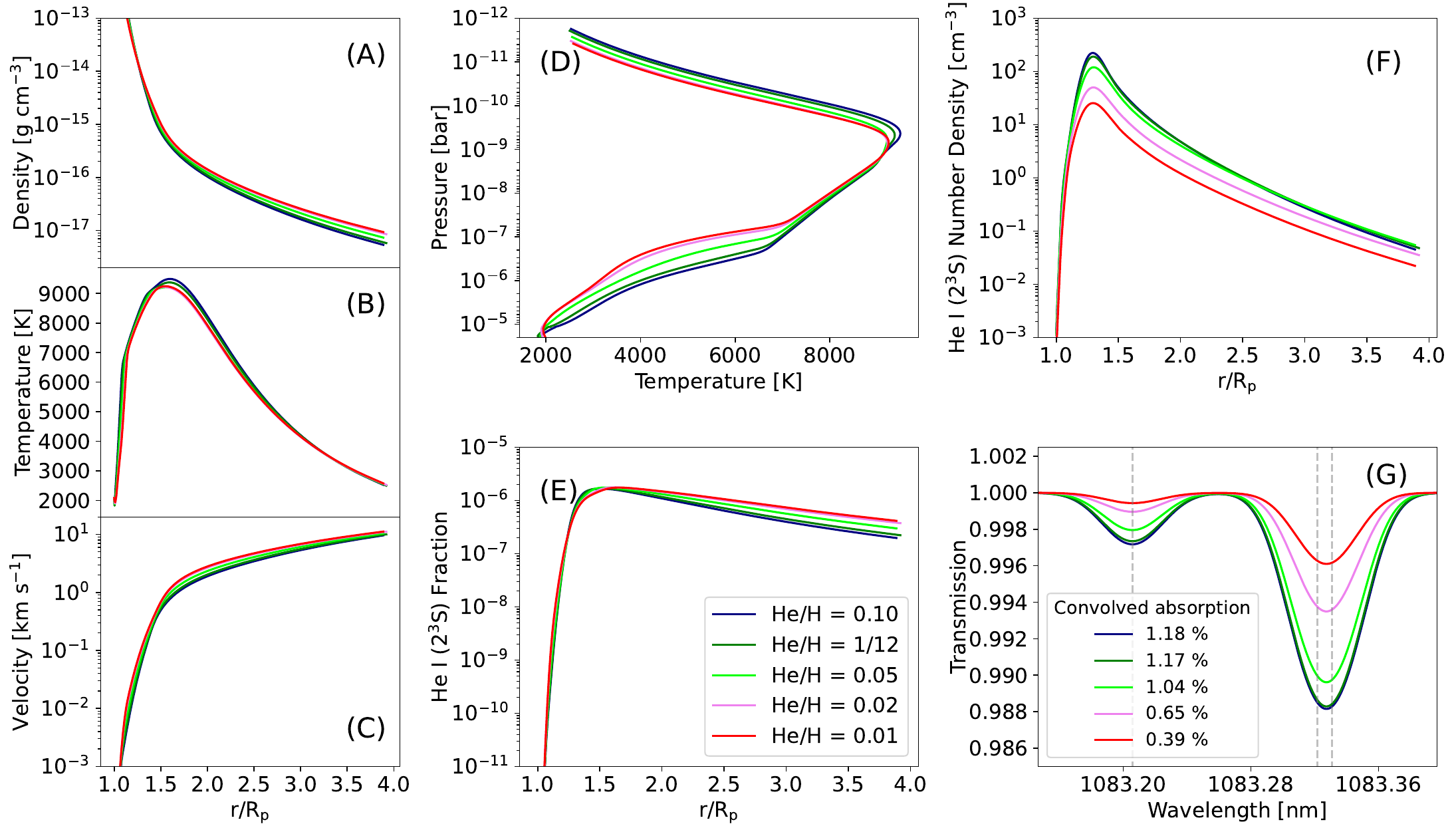}
    \caption{ATES results of KELT-18b simulations with different He/H number density ratio. The outflow density, temperature and velocity are depicted in panels A, B, and C, respectively, for different He/H number density ratio. Panel D shows the pressure-temperature profiles, while panels E and F show the HeI triplet fraction and number density in the planetary outflow respectively. The absorption profiles are shown in panel G with the absorption depth after the convolution with instrument resolution and planetary rotation.}
    \label{fig:ates_result}
\end{figure*}

\section{Mass-loss rate and photoevaporation efficiency maps}
\label{sec:appendix_M_dot_eta}

\begin{figure}[ht!]
   \centering
   \includegraphics[width=0.99\hsize]{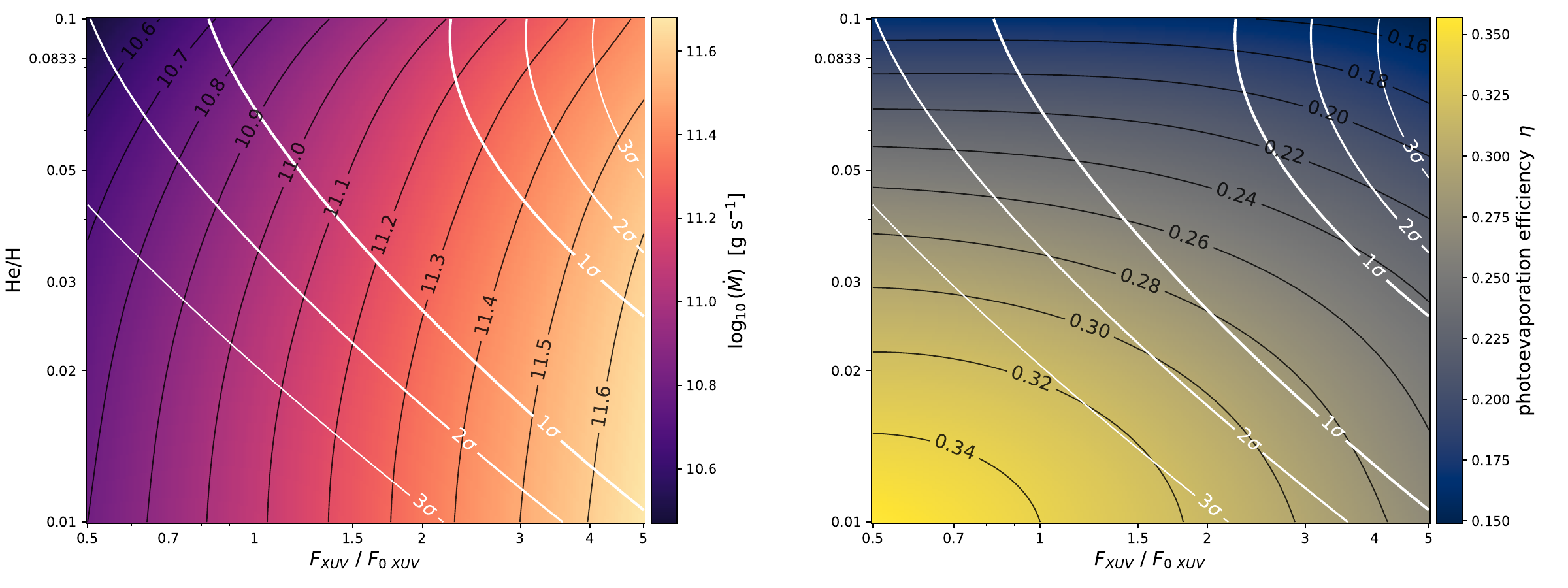} 
   \caption{Maps of the mass-loss rate and photoevaporation efficiency across the explored (He/H, $F_{\text{XUV}}$) parameter space. Left panel: Interpolated map of $\log_{10}(\dot{M}\,[g/s])$. The black lines represent contours of constant mass-loss rate. Right panel: Interpolated map of the photoevaporation efficiency $\eta$, with black lines representing contours of constant efficiency. In both panels, the white curves indicate the $1\sigma$, $2\sigma$, and $3\sigma$ confidence intervals derived from the $\chi^2$ map (Fig.~\ref{fig:chi2_map}).}
   \label{fig:mdot_eta_maps}
\end{figure}

\twocolumn

\section{Supplementary modeling analysis}
\label{sec:appendix_x}

To complement the analysis presented in Sect.~\ref{subsec:quantitative_comparison} and further investigate the observed degeneracies, we performed additional tests on our theoretical modeling to explore the parameter space.

\begin{figure}[ht!]
   \centering
   \includegraphics[width=\hsize]{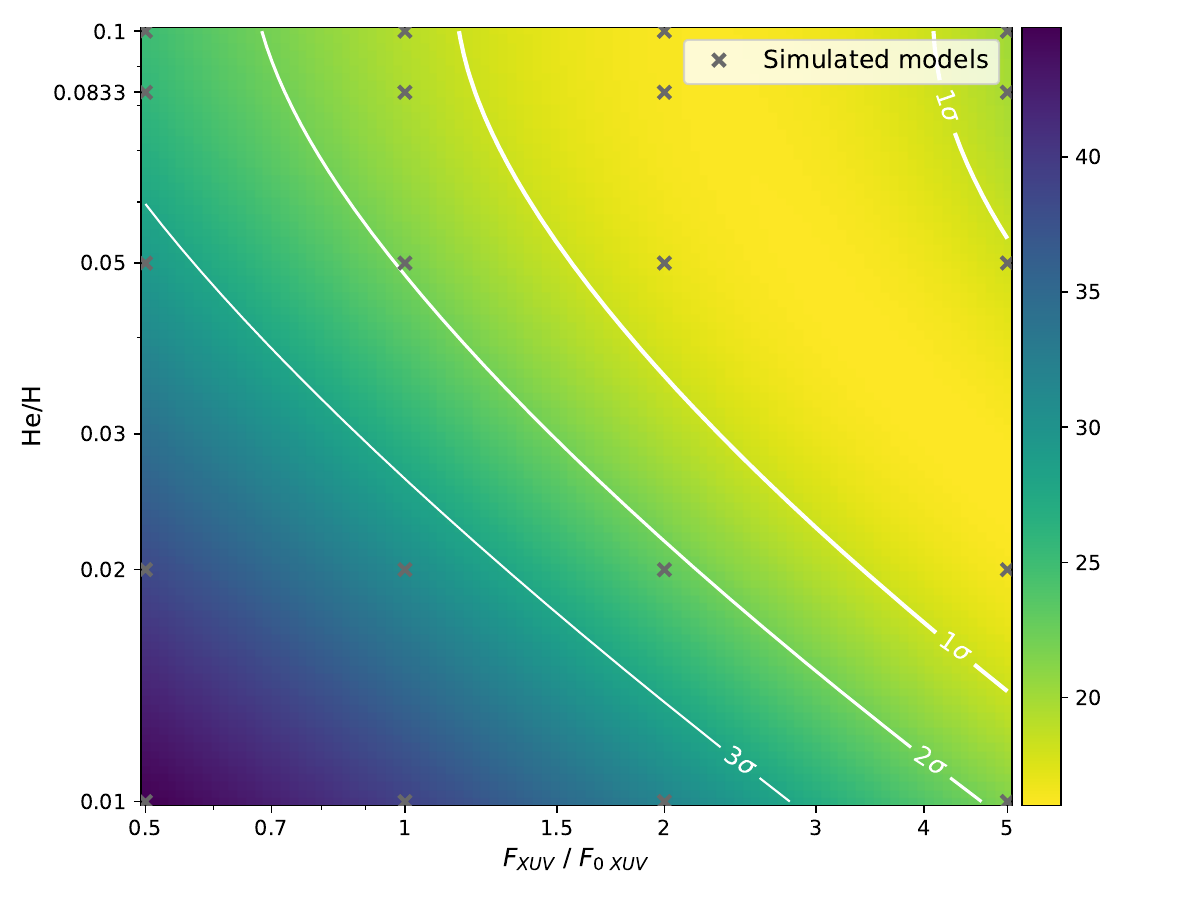}
   \caption{$\chi^2$ map of the explored parameter space including the atmospheric wind broadening. For each point in the (He/H, $F_{\text{XUV}}$) grid, the $\chi^2$ value is minimized with respect to the wind broadening, treated as a nuisance parameter. Contour lines represent the $1\sigma$, $2\sigma$, and $3\sigma$ confidence levels.}
   \label{fig:chi2_map_wind}

\vspace{0.7cm}

   \includegraphics[width=\hsize]{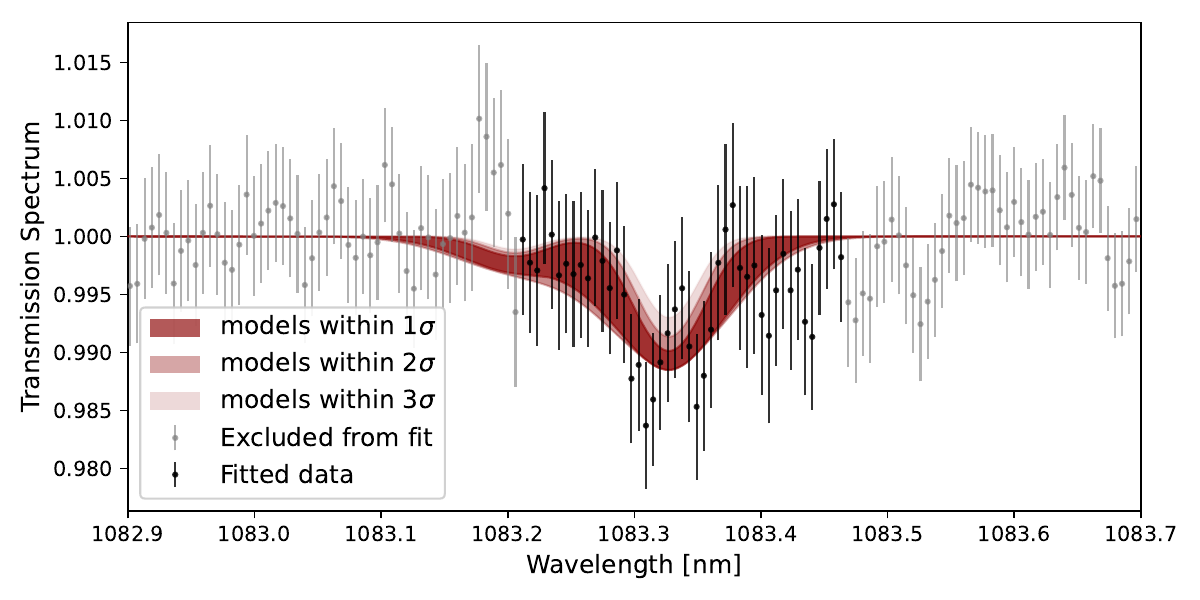}
   \caption{Transmission spectrum of KELT-18b around the metastable He\,I triplet. The shaded regions represent the envelopes of the models compatible at $1\sigma$, $2\sigma$, and $3\sigma$ confidence levels, derived from the $\chi^2$ map that includes the wind broadening (Fig.~\ref{fig:chi2_map_wind}).}
   \label{fig:models_overlapping_wind}
\end{figure}

First, we investigated the impact of atmospheric kinematics on our results by introducing a line-broadening parameter $\sigma_{\text{wind}}$ to account for atmospheric winds. We constructed a new $\chi^2$ map by treating the wind broadening as a nuisance parameter. Specifically, we used a $100 \times 100$ (He/H, $F_{\text{XUV}}$) grid, which is reduced compared to the one adopted in Sect.~\ref{subsec:quantitative_comparison} to maintain computational feasibility. For each grid point, we evaluated 50 equally spaced values for $\sigma_{\text{wind}}$ ranging from 0 to 18\,km\,s$^{-1}$, and retained only the lowest $\chi^2$ value, selecting the wind parameter that best fit the data. The result of this analysis is shown in Fig.~\ref{fig:chi2_map_wind}.

However, given the limited SNR of our transmission spectrum, we are unable to properly resolve the wings of the helium absorption profile, causing the wind parameter to artificially broaden the modeled absorption. Because the wind broadening decreases the line core depth while preserving the equivalent width, the signal is simply redistributed across the unresolved wings. This physical effect has a dual impact on the parameter space. First, models with very high He/H and XUV fluxes can still fit the data without producing an excessively deep core, effectively removing any strict upper limits. Second, the model inherently requires higher He/H and $F_{\text{XUV}}$ values to match the observed absorption depth, shifting the lower significance boundaries upward. This implies that the lower limits derived in the wind-free scenario are conservative, supporting the robustness of our analysis.

\begin{figure}[ht!]
   \centering  
   \includegraphics[width=\hsize]{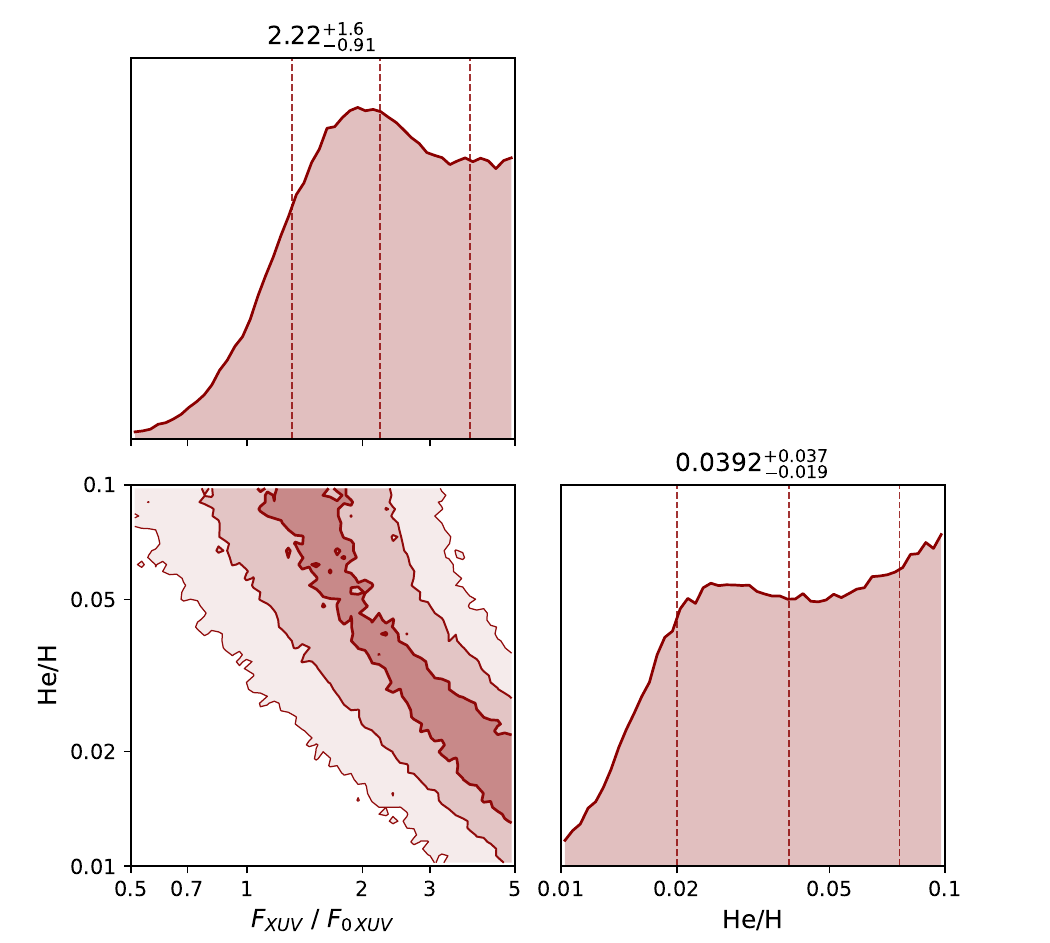}
   \caption{Corner plot showing the 1D and 2D marginalized posterior probability distributions for the He/H and the XUV flux, obtained from the MCMC analysis without the wind parameter. The vertical dashed lines in the 1D histograms indicate the 16th, 50th (median), and 84th percentiles.}
   \label{fig:corner}
\end{figure}

As a second approach, we performed a Markov Chain Monte Carlo (MCMC) fit to test whether an alternative statistical analysis could resolve or improve the observed degeneracy. For the test shown here, we did not include the wind parameter; however, we verified that its inclusion would not alter the final outcome. As shown in the resulting corner plot (Fig.~\ref{fig:corner}), the posterior distributions exhibit the exact same degenerate structure observed in the $\chi^2$ map.
Due to the degeneracy between the parameters, the probability distributions flatten out for increasing values of the XUV flux and He/H. Consequently, the medians and corresponding uncertainties extracted from the 1D marginalized posteriors are not statistically reliable, and they strongly depend on the arbitrary boundaries of the explored parameter space. However, looking at the 2D posterior distribution (He/H vs. $F_{\text{XUV}}/F_{0,\text{XUV}}$), we note that the MCMC confidence intervals are even more stringent than those obtained from the wind-free $\chi^2$ map. This allows us to confirm once again the robustness of the results reported in Sect.~\ref{subsec:quantitative_comparison}.

\end{appendix}
\end{document}